\documentclass[onecolumn]{aa} % for a paper on 1 column  
\usepackage{graphicx}
\usepackage{txfonts}
\begin{document}

   \title{Hierarchical progressive surveys}
   \subtitle{Multi-resolution HEALPix data structures for astronomical images, catalogues, and 3-dimensional data cubes}
   \author{P. Fernique\inst{1} \and M. G. Allen\inst{1} \and T. Boch\inst{1} \and A. Oberto\inst{1} \and F-X. Pineau\inst{1} \and D. Durand\inst{2} \and C. Bot\inst{1} \and L. Cambr\'{e}sy\inst{1} \and S. Derriere\inst{1} \and F. Genova\inst{1} \and F. Bonnarel\inst{1}}
   \institute{Observatoire Astronomique de Strasbourg\thanks{Corresponding authors \email{Mark.Allen@astro.unistra.fr, Pierre.Fernique@astro.unistra.fr}}, Universit\'{e} de Strasbourg, UMR 7550, 11 rue de l'Universit\'{e}, F-67000 Strasbourg, France
% \email{Mark.Allen@astro.unistra.fr}            
   \and
   National Research Council Canada, Canadian Astronomy Data Centre, 5071 W. Saanich Rd., Victoria, BC, Canada}

   \date{Accepted A\&A 19 April, 2015}

% \abstract{}{}{}{}{} 
% 5 {} token are mandatory
 
  \abstract
  {Scientific exploitation of the ever increasing volumes of astronomical data requires efficient and practical methods for data access, visualisation, and analysis. Hierarchical sky tessellation techniques enable a multi-resolution approach to organising data on angular scales from the full sky down to the individual image pixels.}
  {We aim to show that the Hierarchical progressive survey (HiPS) scheme for describing astronomical images, source catalogues, and three-dimensional data cubes is a practical solution to managing large volumes of heterogeneous data and that it enables a new level of scientific interoperability across large collections of data of these different data types.}
   {HiPS uses the HEALPix tessellation of the sphere to define a hierarchical tile and pixel structure to describe and organise astronomical data. HiPS is designed to conserve the scientific properties of the data alongside both visualisation considerations and emphasis on the ease of implementation. We describe the development of HiPS to manage a large number of diverse image surveys, as well as the extension of hierarchical image systems to cube and catalogue data. We demonstrate the interoperability of HiPS and Multi-Order Coverage (MOC) maps and highlight the HiPS mechanism to provide links to the original data.}
   {Hierarchical progressive surveys have been generated by various data centres and groups for $\sim$200 data collections including many wide area sky surveys, and archives of pointed observations. These can be accessed and visualised in Aladin, Aladin Lite, and other applications. HiPS provides a basis for further innovations in the use of hierarchical data structures to facilitate the description and statistical analysis of large astronomical data sets.}
   {}


   \maketitle
   
%
%________________________________________________________________

\section{Introduction}

Astronomical surveys cover the sky in a rich tapestry of data as the recorded signals of photons with spectral and temporal characteristics, stitched together into multi-dimensional maps for science. Large surveys may cover the entire sky, others span irregular patchwork patterns of coverage, and the collections of pointed observations in observatory archives typically project sparse sky coverage maps composed of thousands of instrumental fields of view. Systems for managing and manipulating astronomical data that cover this wide range of spatial scales from the individual pixels up to the full sky, which may also have spectral or temporal axes, are necessary for astronomical research that uses data from many surveys and archives. The sheer multitude of surveys and archived data requires that these systems are efficient, interoperable, and easy to implement and use. 

Current and planned astronomical projects are ushering astronomy into an era of petabyte surveys, where source catalogues may  contain 10$^{9}$-10$^{12}$ entries.  To highlight a few examples: The Gaia mission \citep{2001A&A...369..339P} is taking data and will measure the positions of about one billion stars; LOFAR \citep{2013A&A...556A...2V} science data products are expected to grow at five petabytes\,yr$^{-1}$ and will generate petabyte size catalogues;  the Euclid\footnote{\url{http://sci.esa.int/euclid/}} mission complete survey will comprise hundreds of thousands of images (several tens of petabytes of data), and the expected catalogues will have some ten billion sources. LSST \citep{Ivezic:2008ub} will revolutionise the exploration of the transient sky by accumulating some 32$\times10^{12}$ photometric measurements, and the SKA\footnote{\url{https://www.skatelescope.org}} is planning surveys with exabyte volumes. Scientific exploitation of these rich surveys, in particular the pursuit of research that is based on their combination and comparison, requires that these data are highly interoperable and that the systems that enable access to these data be tightly integrated with the whole process of data collection, processing, curation and archiving, and analysis.

\begin{figure*}
\includegraphics[width=17cm]{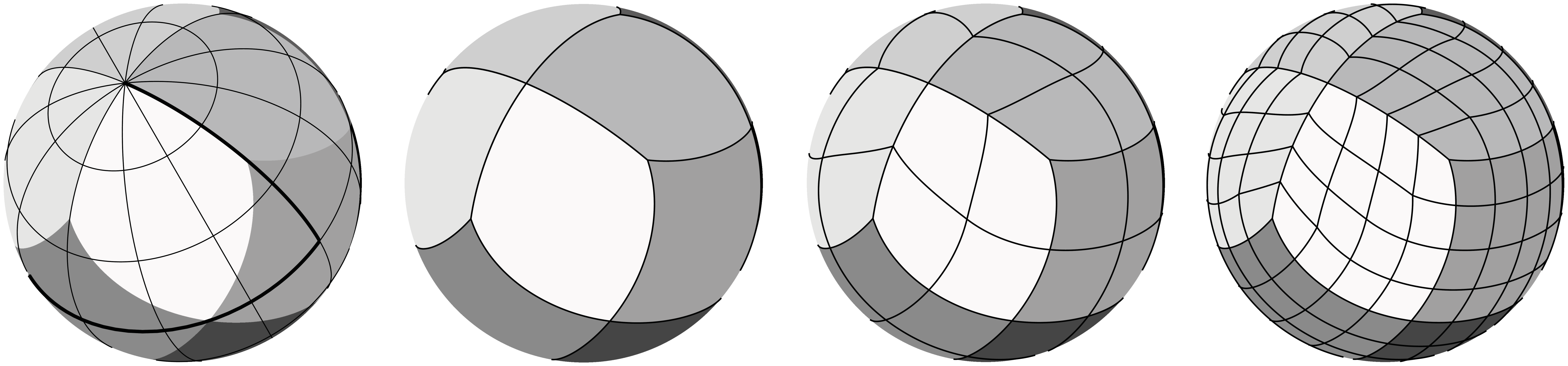}
\caption{HEALPix tessellation of the sphere. Each of the four orthographic projections of the sphere shown here display the twelve base resolution HEALPix quadrilaterals shaded with different grey levels. The left most frame shows how the HEALPix tessellation relates to a generic spherical longitude and latitude coordinate system. The second, third, and forth spheres to the right are overlaid with HEALPix grids of orders $k=0,1,2$ respectively, to illustrate the hierarchical structure of HEALPix where each pixel is divided into four self-similar pixels at each successive order. 
}
\label{spheres_fig}
\end{figure*}

There are already hundreds of large surveys available though astronomy data centres and services. It is also becoming common to consider the observational archives of the pointed observations of a given instrument as a kind of heterogeneous survey data.  This is also true of the source catalogues derived from cumulated archives of pointed observations, for example the Chandra source catalogue \citep{2010ApJS..189...37E} and the Hubble source catalogue \citep{2012ApJ...761..188B}.  With increasingly interoperable data in observational archives, an ultimate goal could involve characterising observational data along many physical axes so that the data may be described within a generalised multi-dimensional parameter space, and data could be combined along any compatible axis. Presently, there is growing awareness of the potential of practical schemes for characterising the spatial sky coverage of the wide diversity of astronomical survey and pointed observation archival data. Our position is that characterising the spatial coverage of this diversity of data using a hierarchical scheme leads to innovative solutions to accessing, visualising, and manipulating the data, and allows for general cross-comparison of data that scales to many hundreds of data sets.

Interoperability of astronomical image data is already very good due to compatible formats (FITS) and the use of standardised astronomical coordinate systems \citep{2010A&A...524A..42P}. Many surveys and archives are made available through web interfaces with various kinds of database systems, and virtual observatory (VO\footnote{\url{http://www.ivoa.net}}) metadata and protocols are helping to standardise these interfaces, leading to services that provide coordinated access to multiple surveys and data sets. These interfaces typically rely on queries expressed in astronomical coordinates and use a \emph{query, select, download} pattern for accessing cut-out regions of the data. For large area studies this approach can be cumbersome because downloading the data for large sky regions at the original pixel resolution is inefficient and it is difficult to use a survey as a whole, let alone multiple surveys at the same time.  Visualisation interfaces have however made significant progress towards interactive displays of multiple surveys across the entire sky. The hierarchical techniques that have been employed to enable fast multi-resolution visualisation of astronomy data can be extended to enable easy publication and access to image survey data.  These techniques can also be generalised to describe heterogeneous archival data, multi-dimensional cube data, as well as catalogue data, and have the potential to provide a new level of interoperability between these different data types.

Over the past ten years there has been much improvement in the tools used for the visualisation of astronomical data \citep{2011PASA...28..150H}.  For large astronomical imaging surveys, it is now commonplace to browse, pan, and zoom into the survey data using a tool that accesses the data over the internet and streams only the portions of the data that are needed for the current user view.  This kind of remote visualisation enables the exploration of large data sets from the full wide scale view of an entire survey displayed on a grid of the whole sky, down to the detailed zoomed-in view at the finest spatial limits of the images. Examples of such interactive interfaces include Google sky\footnote{\url{http://www.google.com/sky}}, Microsoft World Wide Telescope\footnote{\url{http://www.worldwidetelescope.org}} (WWT),  Sky-Map/Wikisky\footnote{\url{http://sky-map.org}} and Aladin\footnote{\url{http://aladin.unistra.fr}}.  These tools and interfaces are made available as client software, in-browser tools, and most recently as applications for mobile devices. These interfaces serve a range of visualisation purposes for professional research astronomy, as well as education, citizen science, and public outreach, with their ease of use being a major factor in their up-take. 

All of the sky browsing tools mentioned above must balance the considerations of internet bandwidth and the performance of the client displays so that the real-time update of the display for panning and zooming is fast enough to be interactive. Important considerations in the design of systems for interactive visualisation of remotely stored data are the underlying organisation of the data, and how the sky is divided up into regions.  All of the applications mentioned above employ a hierarchical multi-resolution tessellation of the sky. Several different schemes are in use:  Aladin uses the Hierarchical Equal Area isoLatitude Pixelization scheme, HEALPix\footnote{\url{http://healpix.sourceforge.net}}  \citep{2005ApJ...622..759G, 2011ascl.soft07018G}; WWT uses a hierarchical triangular mesh (HTM) \citep{2001misk.conf..631K} and an extension of this system called tessellated octahedral adaptive subdivision transform (TOAST); Google sky uses a cylindrical method. These different tessellations of the sky have been developed for different purposes and some aspects of their relative merits are discussed in \citet{2001misk.conf..638O}. For sky browsing applications, the choice of tessellation scheme dictates how the original data must be mapped onto a set of hierarchically organised tiles, and determines the spatial index to be used in the applications.

This paper describes the Hierarchical progressive survey (HiPS) system that has been designed to facilitate the access and visualisation of astronomical survey data. HiPS is based on the HEALPix sky tessellation, and is essentially a mapping of survey data at various spatial resolutions into a collection of HEALPix tiles.  HiPS has been developed at the Centre de donn\'{e}es astronomiques de Strasbourg (CDS\footnote{\url{http://cds.unistra.fr}}) to support the visualisation of all sky imaging survey data in Aladin \citep{2010ASPC..434..163F}, and has been extended to also support source catalogues and three-dimensional cube data. HiPS is designed specifically for astronomical data in that it is fundamentally spherical, it takes the astrometric and photometric properties of the original data into account, and emphasis is placed on ease of use with no need for special servers or database systems. Section \ref{HiPS_section} of this paper specifies the hierarchical spatial index scheme and its implementation as a file system directory and file structure. Section \ref{MOC_section} describes the relationship between HiPS and the Multi-Order Coverage \citep[MOC;][]{Boch:2014vv} maps that define regions of the sky.  Section \ref{Generation_section} describes how HiPS is applied to astronomical images, catalogues, and three-dimensional data cubes. Section \ref{HiPS_avail_section} highlights the $\sim$200 HiPS data sets that are currently available from various astronomy data centres and outlines how these can be used via the Aladin, Aladin Lite, and other compatible client applications. In Section \ref{Discussion_section} we discuss future directions for the use of hierarchical data structures to visualise and analyse astronomy data.

\section{Hierarchical progressive surveys}
\label{HiPS_section}

\subsection{HEALPix}

HEALPix is a curvilinear partitioning of the sphere that supports a hierarchical tree structure for multi-resolution applications. The detailed geometry and properties of HEALPix are described in \citet{2005ApJ...622..759G}. Here we summarise the properties that are necessary for the definition of HiPS.

The HEALPix partitioning of the sphere uses a base resolution that divides the sphere into 12 quadrilateral pixels, which are recursively divided into four self-similar smaller pixels in a 2$\times$2 pattern to create finer and finer higher order meshes (see Figure~\ref{spheres_fig}).  The resolution of the mesh at a given order $k$ is defined by $N_{side}$, which is the number of divisions of the base resolution pixels and which doubles at each successive order, so that a HEALPix map of $N_{side}$=2$^{k}$ consists of 12$N_{side}^{2}$ pixels. All of the pixels at a given order have equal area of $\Omega_{pix}=\pi/(3N_{side}^{2})$, and the pixel centres are arranged in rings of equal latitude. HiPS uses the HEALPix nested numbering scheme \citep{2005ApJ...622..759G}.

\subsection{HiPS tiles and pixels}
 
HiPS makes use of the hierarchical features of HEALPix to organise astronomical data into HEALPix maps of different orders.  In this section we describe the HiPS tile and pixel structure for image data where both the tiles and their pixels are defined by HEALPix orders. This provides the general framework for HiPS that is later applied to other types of data. 

For image surveys we start by resampling the images onto a set of HEALPix maps of increasing order $k$. Each HEALPix map generated in this way is a representation of the original image data mapped onto a HEALPix mesh with a different pixel size, and each successive order has four times the number of pixels of the previous order. This multi-resolution representation of the original images provides the basis for visualising the data in a progressive way as the pixels that are required for a given view can be accessed from the pre-computed HEALPix maps, and the nested pixel numbering scheme provides a simple hierarchical indexing system that encodes pixel inheritance across the different orders.  

\begin{figure*}
\includegraphics[width=17cm]{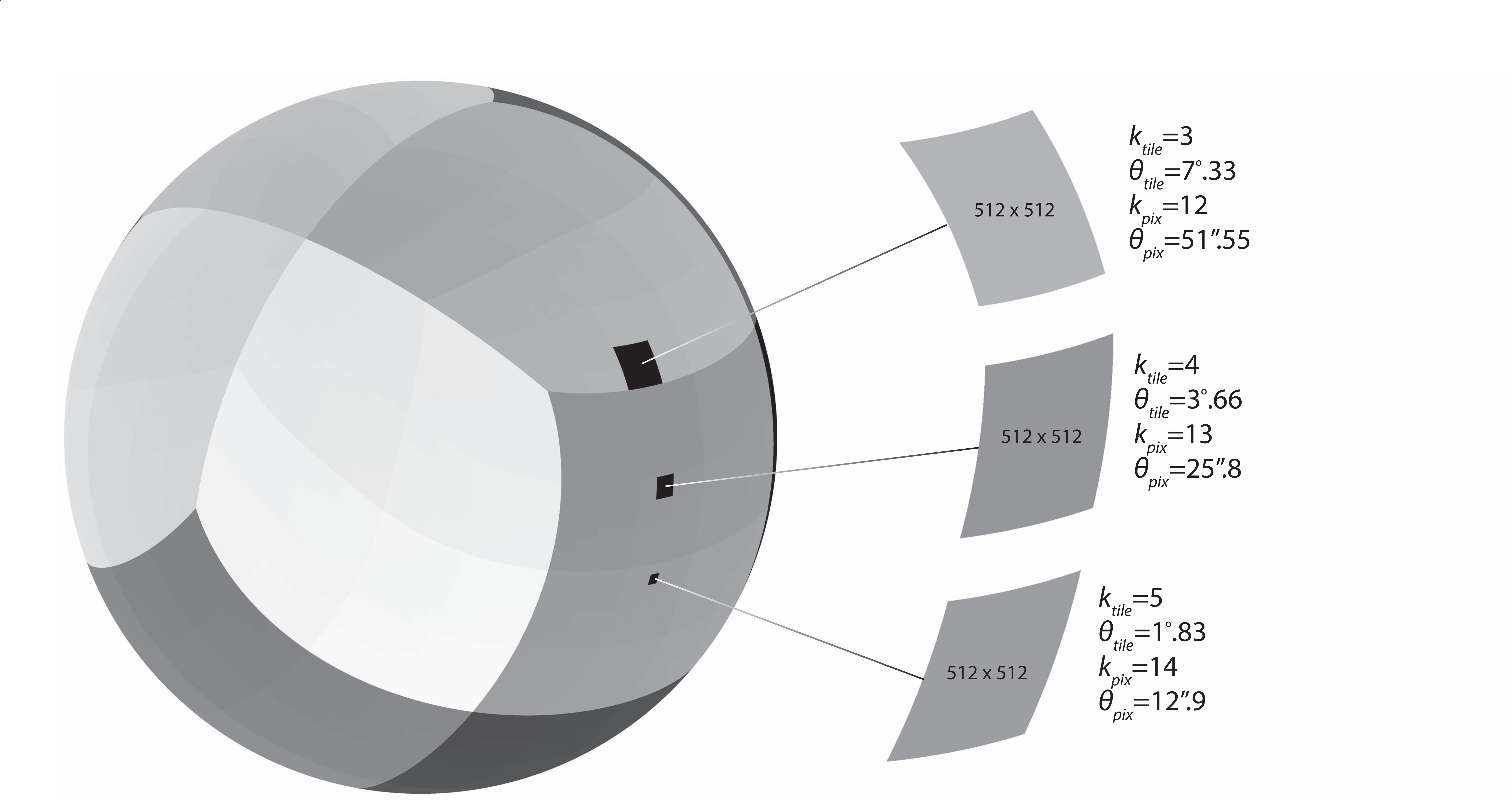}
\caption{HiPS tiles and pixels. 
A HiPS structure is defined by a set of tile orders $k_{tile}$ and by the number of pixels along each side of the tiles $n_{tile}$. Here we show an orthographic projection of the sphere shaded to indicate different orders, and highlighting in black three individual HiPS tiles of orders $k_{tile}$=3,4 and 5, with pixel tiling of $n_{tile}$=512.  Note that HiPS representations of astronomical data usually start with $k_{tile}$=3 as the lowest order tile, and each successive order of smaller tiles is a higher resolution representation of the data. Some characteristics of the tiles are shown in the expanded view of the tiles on the right of the figure, including the approximate side length of the tile $\theta_{tile}$ (defined as the square root of its area $\theta_{tile}$ = $\Omega_{tile}^{1/2}$). The number of pixels inside a tile is constant over all of the orders, and in this example where $n_{tile}$=512  each tile comprises 512$\times$512 pixels. The pixels inside a tile correspond to a higher order HEALPix grid, and in this example $k_{pix}$=$k_{tile}$+ $\log_2n_{tile}$=$k_{tile}$+ 9, so that the $k_{tile}$=3 tiles are filled with 51\farcs55 pixels that correspond to a HEALPix grid of order 12.}
\label{Hips_fig}
\end{figure*}

It would however be impractical to access and transfer individual HEALPix pixels, so the HiPS scheme groups pixels into tiles as the basic unit of a map that is transferred when any of its pixels are needed. The specification of the tiles makes use of the hierarchical properties of HEALPix to define the tiles as the HEALPix pixels of a lower order HEALPix grid, so that both the tiles and the pixels within the tiles are described by the HEALPix geometry. The general relationship between the tiles and pixels is that a tile with  
$n_{tile}$ pixels along each side forms a HEALPix mesh of order of $k_{tile}<k$ such that $\log_2(n_{tile})= \log_2(N_{side}/N_{side,tile})=k - k_{tile}$.  The HiPS structure is therefore in general defined as a set of HEALPix maps with orders $k$ and tiling $n_{tile}$. The HiPS tiles and pixels are illustrated in detail in Figure~\ref{Hips_fig}.

For astronomical purposes the generic HEALPix spherical coordinate system \citep[$\theta, \phi$ as in][]{2005ApJ...622..759G} is mapped to astronomical coordinates so that the positions of the pixels have precisely defined positions on the sky (Figure~\ref{spheres_fig}). For HiPS we consider only two astronomical coordinate systems; the equatorial coordinate system defined by the international celestial reference system (ICRS), and also the Galactic coordinate system. Limiting the coordinate system to these two options is intended to provide a good level of consistency while also accounting for current use of HEALPix. While there has been significant use of HEALPix with the galactic coordinate system for cosmic microwave background (CMB) data, it is much more common for survey data in other wavebands to be described by equatorial coordinates.  ICRS is chosen as the default coordinate system for HiPS with Galactic coordinates as the alternative option.

The HiPS representation of a collection of astronomical images is comprised of a set of HEALPix maps over a range of successive orders $k_{min}$ to $k_{max}$.  The choice of the number of pixels in the tiles is based on the need for efficient transfer of the tiles to an interactive visualisation client. For HiPS that only use low orders with original data resolutions \la1\degr, this is not so important because the volume of data to be transferred is never very large,  and these data are best served by tile sizes $n_{tile}$=64 to 256. For higher resolution data, practical tests with bandwidth of $\sim$1\ Mbit\ s$^{-1}$ shows that tiles of 512$\times$512 pixels (i.e. $n_{tile}=512$) are a good balance between the size of the files needed to store the arrays, and the number of pixels to be displayed and updated in the (typically 1000$\times$1000 screen pixels) window of a visualisation client tool. 

The properties of HiPS map tiling and the relationship of the tiles to HEALPix pixels are shown in Table \ref{table:1}. The first four columns of this table show the properties of HEALPix; $k$, $N_{side}$, $N_{pix}$, and $\theta_{pix}$ following Table 1 of \citet{2005ApJ...622..759G} which has been extended here to show all orders up to $k$=29. As described above, the HiPS tiles that enclose the pixels of a given HEALPix order are defined by a lower order HEALPix grid, and this is illustrated in the table with our commonly used case of $n_{tile}$=512. The three next columns thus show; the HEALPix order of the tiles $k_{tile,512}$, the number of such tiles that cover the sky $N_{tile,512}$, and the size of these tiles $\theta_{tile,512}$.  The zeroth order grid for the $n_{tile}$=512 HiPS ($k_{tile,512}$=0) is therefore comprised of 12 tiles of size $\sim$58\fdg6, where the 512$\times$512 pixels of the tiles correspond to HEALPix pixels of order $k$=9. Note that the nine order difference between the HiPS tile grids and the HEALPix pixel grids means that the $k_{tile,512}$, $N_{tile,512}$, and $\theta_{tile,512}$ columns are shifted versions of the $k$, $N_{pix}$, and $\theta_{pix}$ columns.  Use of different choices for $n_{tile}$ would result in a different shift, for example for some low resolution data sets we sometimes employ $n_{tile}$=64, for which the values of $k_{tile,64}$, $N_{tile,64}$, and $\theta_{tile,64}$ would correspond to a shift of $\log_2n_{tile}$=6 rows in the table with respect to the $k$, $N_{side}$, and $N_{pix}$ columns. The properties of single order HiPS with $n_{tile}$=64, 128, and 256 are shown in Table~2.

\subsection{HiPS file structure and indexing}

The tiles define the basic unit of storage for the set of HEALPix maps that make up a HiPS.  Due to their quadrilateral nature, the HEALPix tiles are readily encoded as square arrays that can be stored in a number of different file formats. The simplification of having square arrays is a major advantage of HiPS as it leads to straight forward methods for drawing the tiles in any sky projection.  It also means that the description of the array stored in the file is very simple in that all of the array positions are filled.

The HEALPix indexing scheme provides a tree structured numbering system for HEALPix pixels with a simple relationship between the pixel numbers at different orders. Using the nested system leads to a simple organisation of the tiles into a filesystem directory structure. Each tile is encoded as a file, and all of the tiles for a given order are stored in a directory for that order. Starting at the third order ($k_{tile,512}$=3), the directories are named Norder3, Norder4... Norder$k_{tile,512}$ up to the maximum tile order. Within each Norder directory\footnote{The Npix files benefit from being grouped into further sub-directories of 10000 Npix files named Dir10000, Dir20000, etc. In this scheme, an individual pixel $n$ at order $k$ is in the tile $m=n/(2^{10})$ at order $k_{tile,512}=k-9$, and will be found in the file Norder$(k-9)$/Dir$(m, 10000)$/Npix$m$.} the tiles are named according to their tile number within that order i.e. Npix0, Npix1... Npix$(N_{tile,512}-1)$. 

The square arrays of pixel values in the tile files are arranged so that the pixels are in their natural spatial order. This means that the array when displayed as a square image, is a slightly distorted view of the curvilinear quadrilateral area of the sky. The natural spatial ordering of pixels in the tile means that no pixel indices or astrometry information needs to be stored in the file, and the simple ordering also facilitates the computation of the geometric projection of the pixels onto a display of the sky (see section \ref{Tools_section}). Also, the use of the simple files and directories (compared to a database management system, or special format) means that HiPS are easily transportable and may be made available over the internet without the need for special servers or protocols. HiPS files may be simply put online with a standard web (HTTP) server.

 \begin{figure*}
\includegraphics[width=17cm]{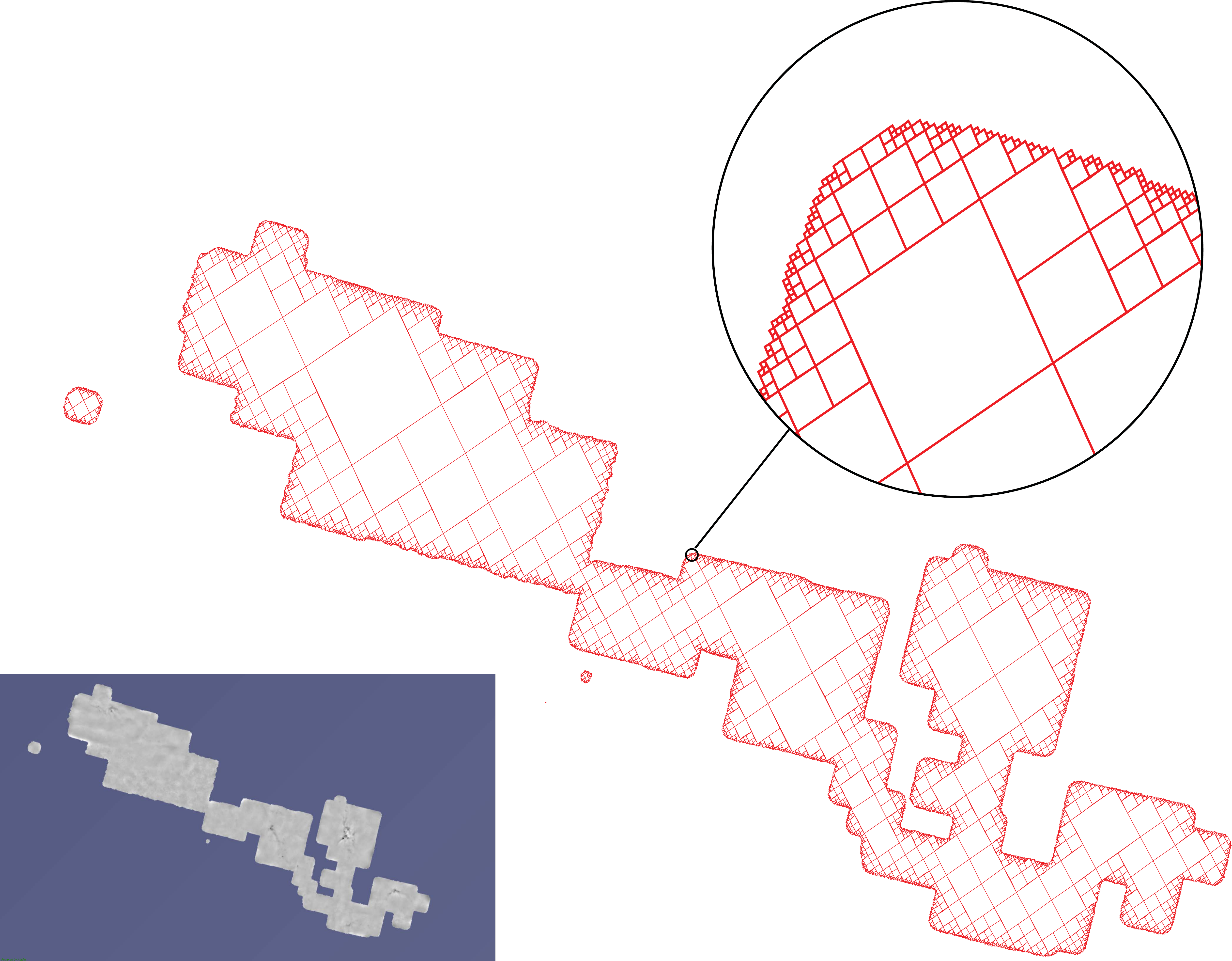}
\caption{Multi-Order Coverage (MOC) map of the SCUBA 850\ $\mu$m data. The lower left panel shows an approximately 5\degr$\times$2\fdg5\ region of the HiPS image of the SCUBA 850um map. The MOC map of this region is shown in red, where we display the individual HEALPix pixels of different orders that define the MOC.  The lowest order (largest HEALPix pixel) in this MOC is $k$=6  (corresponding to $\theta$=55\farcm0), and the range of orders extends up to $k$=15 as shown in the magnified section of the figure where the smallest ($\theta$=6\farcs44) pixels define the resolution of the MOC.}
\label{MOC_fig}
\end{figure*}

\section{Multi-order coverage maps}
\label{MOC_section}

The HEALPix framework has also been used as the basis for the hierarchical multi-order description of sky regions called Multi-Order Coverage (MOC) maps. The MOC system exploits the same hierarchical characteristics of HEALPix as used by HiPS, in this case to describe regions of the sky in terms of coverage by HEALPix pixels of different orders. MOC has already been accepted as an IVOA recommendation \citep{Boch:2014vv} for the description of sky regions. The common underlying framework of HEALPix means that MOC and HiPS are both highly interoperable and complementary to each other.

MOC maps can support the description of many types of sky coverage from a complete 4$\pi$ sr full sky survey, to the non-contiguous complex coverage of a set of pointed observations. The multi-order nature of MOC maps allows them to be defined at any HEALPix resolution to support different kinds of usage. Coverage maps that characterise the outlines of sky regions are very useful for many purposes such as establishing intersections, unions, and other logical operations based on the geometric coverage of regions of the sky. 

MOC maps are defined by first considering the set of HEALPix pixels at a given order that cover a region on the sky. This set of pixels is simply described as a list of HEALPix pixel indices in the nested numbering scheme.  The self-similar hierarchical properties of HEALPix can then be used to compact the list into a concise and unique set of HEALPix pixel indices from multiple HEALPix pixel orders that provide the equivalent sky region.  As such, MOC maps can be directly defined from any kind of data that is distributed on the sky.

In practice a MOC map is constructed by selecting a maximum HEALPix order that will determine the size of the smallest HEALPix pixels to be used, and then listing all of the pixels that intersect the data. The list is then compressed by replacing any four consecutive tiles at order $k$ by their parent tile from order $k-1$. This is done recursively down to $k=0$, resulting in a list of HEALPix indices from different orders. The encoding of these lists of HEALPix indices into various formats including ASCII, and their serialisation into the FITS format, makes use of NUNIQ packing as described in \cite{Boch:2014vv}.

A MOC description of a sky region is an approximation of the exact region boundaries. The resolution of a MOC is determined by the highest order used, and the outline is always limited to the compound shape of multi-order HEALPix pixels. Higher order MOC maps can be used to improve accuracy, but the limitations of the MOC approximation must be taken into account when using it for different purposes. While detailed schemes for expressing the exact coverage of sky regions are available, e.g. \citet{Rots:2011ul}, the emphasis of MOC is on facilitating its ease of use and implementation. 
Another benefit of the simplicity of MOC is that the description of a sky region is unique, whereas schemes that use geometric constructions in various coordinates systems can be described in multiple (non-unique) ways especially in the case of complex regions.
 
Figure~\ref{MOC_fig} provides an example of a MOC that represents the coverage of a portion of the SCUBA (Submillimetre Common User Bolometer Array) 850$\mu$m data (see P/SCUBA/850em in Table 3.). SCUBA is a bolometer camera on the James Clerk Maxwell telescope (JCMT) \citep{1999MNRAS.303..659H}. A HiPS image of all of the public 850 $\mu$m observations has been constructed by the CADC. The combined data have an irregular pattern of coverage on the sky because of the distribution of the telescope pointings and the scanning characteristics of the array. The MOC representation of this data spans all of the regions where the data is defined, and provides a convenient footprint of the survey region on the sky. This MOC has been calculated by identifying all of the HEALPix pixels of order $k=9$ (corresponding to a resolution of 6\farcm871) which include valid data, and then compacting this list of pixel indices to create a unique list of HEALPix pixels over all orders $k$ \lid\ 9. The resulting MOC displayed in Figure~\ref{MOC_fig} shows how the HEALPix pixels of different orders fill out this complex and non-contiguous pattern of coverage on the sky. Compacting the pixels allows the larger contiguous areas to be efficiently filled with lower order pixels, while the finer details of the shape are represented by the smaller pixels at the chosen resolution.

\section{HiPS for images, source catalogues, and three-dimensional data cubes}
\label{Generation_section}

The development of the HiPS scheme has been driven by the need for scientifically robust hierarchical access to image survey data. The most direct use of HiPS is thus the mapping of astronomical images onto HiPS tiles. The HiPS scheme is however not limited to images, rather it can be used for many kinds of data. The concept that promotes the use of HiPS beyond its application to images is that the HiPS tiles may be used as general \emph{containers} for any sort of information that is related to the sky coverage of the tiles. The content of the tile container can then vary according to different objectives. HiPS data structures have so far been successfully generated for astronomical source catalogues and multi-dimensional cube data, and the tile container has also been found to be very useful for storing links to the detailed metadata about the original progenitor data associated with the tiles.

Here we describe the considerations to be taken into account when generating HiPS for different types of data; images, catalogues, and three-dimensional data cubes.

\subsection{Images}

As described in Section~\ref{HiPS_section}, the HiPS representation of an image survey is generally constructed by resampling the images onto a HEALPix grid at the maximum desired order $k_{max}$, and then generating the tile images for each of the tile orders. The process for resampling the pixel values from the original images depends on the intended purpose, and involves all of the same issues as generally encountered when mosaicking images. In general it is necessary to consider the desired angular resolution and resampling of the images, and also the choice of the methods to be used for combining data in regions where images overlap, and how to deal with variations in the background level. 

The choice of the highest order $k_{max}$ for a HiPS determines the minimum pixel size (see Table~\ref{table:1}) and is usually chosen to be close to the angular pixel size or resolution of the original data. To allow for various potential uses of HiPS no restrictions are made on the algorithms for resampling the original data onto the pixels, nor on the methods used for construction of the hierarchy of orders. The HiPS that are currently available for a wide range of surveys and image data sets demonstrate the benefits of some of the different approaches that have been tried so far. Many of these efforts have focused on the display quality of the final results, and others have placed greater emphasis on photometric accuracy and reducing re-sampling effects. These issues also guide the choice of file format to be used because of the different display efficiencies and dynamic ranges that may be supported by different file formats. 

The methods for generation of successive orders $k_{max}$-1 to $k_{min}$ of HiPS hierarchies may also be freely chosen depending on the intended purpose. The self-similar nature of the tiles means that conservation of flux in order $k$ is achieved by simple averaging of each set of 2$\times$2 sub pixels in order $k-1$.  The desired dynamic range of the pixel values influences the choice of the image file format. HiPS tiles are most appropriately stored as FITS format files when the full original dynamic range of the data must be preserved. For better performance however, JPEG or PNG encoding of the tiles is more appropriate. In this case with reduced dynamic range it is better to construct the successive HiPS orders using the median because it more reliably preserves the interesting structure in the images. For most cases we have chosen to generate both FITS and PNG format files.

Construction of HiPS also requires consideration of the choice of the lowest order to include. Low order tiles ($k_{tile}$=0,1, and 2) correspond to large areas of the sky with tile side lengths $\theta_{tile}>10\deg$. These orders are typically not useful for display of astronomical images because they are too coarse even for a full sky projection. Accordingly HiPS maps usually start at order $k_{tile}$=3 and include all of the successive orders up to a maximum order chosen so that the HEALPix pixel size is close to the original image pixel size.  

Our construction of a HiPS representation of the well known Digitized Sky Survey\footnote{The digitized sky surveys were produced at the Space Telescope Science Institute under U.S. Government grant NAG W-2166. The images of these surveys are based on photographic data obtained using the Oschin Schmidt Telescope on Palomar Mountain and the UK Schmidt Telescope. The plates were processed into the present compressed digital form with the permission of these institutions. See \url{https://archive.stsci.edu/dss/faq.html}} 
(DSS2) illustrates a number general aspects about creating HiPS for all-sky surveys. The digitized photgraphic plate images of the DSS2 ($\sim$1.5 million FITS format images of 768$\times$768 pixels) have an original pixel scale of $1\farcs0$\ pixel$^{-1}$ and cover the entire sky. To map the DSS2 onto HiPS tiles of different orders we first identify the HEALPix pixel order that corresponds to the resolution of the survey, in this case $k$=18 with $\theta_{pix}$=0\farcs81 is a good choice as it slightly over-samples the 1\farcs0 DSS2 pixels. A HEALPix map of the whole sky at this order contains $N_{pix}$=8.25$\times10^{11}$ pixels. Choosing the HiPS tiles to be of our preferred size of 512$\times$512 pixels means that the highest tile order will be $k_{tile,512}$=9, corresponding to a tile size of $\theta_{tile,512}$=6\farcm87. 

The most important step in the construction of the HiPS is the mapping of the DSS2 pixels onto the set of 0\farcs81  HEALPix pixels of order $k$=18. To re-sample the DSS2 pixel values onto the HEALPix pixels we use a bi-linear interpolation of the four closest DSS2 pixel values. When multiple overlapping input images contribute to a pixel we find that applying weights to input pixel fluxes as a function of the distance to the closest edge provides good results for the display quality of the tile images. (As the DSS2 pixel values are a measure of the photographic density of the original plate, which is non-linear with the intensity, we concentrate on display quality rather than photometric considerations in this example). Once the full set of $k$=18 pixels has been computed from the original DSS2 pixels it is straight forward to construct all of the lower orders by taking the median of each 2$\times$2 pixels to create the successive orders. At each order we group the pixels into 512$\times$512 tiles so that the tiles and pixels of the different orders have properties as shown in Table \ref{table:1}.  The lowest tile order included is $k_{tile}$=3 ($\theta_{tile,512}$=7\fdg33), which has pixels of HEALPix order $k$=12 ($\theta_{pix}$=51\farcs5). 
 
The resulting HiPS of the DSS2 can be visualised on all scales from the whole sky down to the resolution of the survey by streaming the 512$\times$512 pixel tiles of the required orders to a tool that can draw those tiles on a projection of the sky. Zooming and panning of the HiPS simply requires knowledge of the HiPS indexing structure so that the correct tiles for a given view can be accessed and mapped onto the display. 
 
\begin{figure*}
\includegraphics[width=17cm]{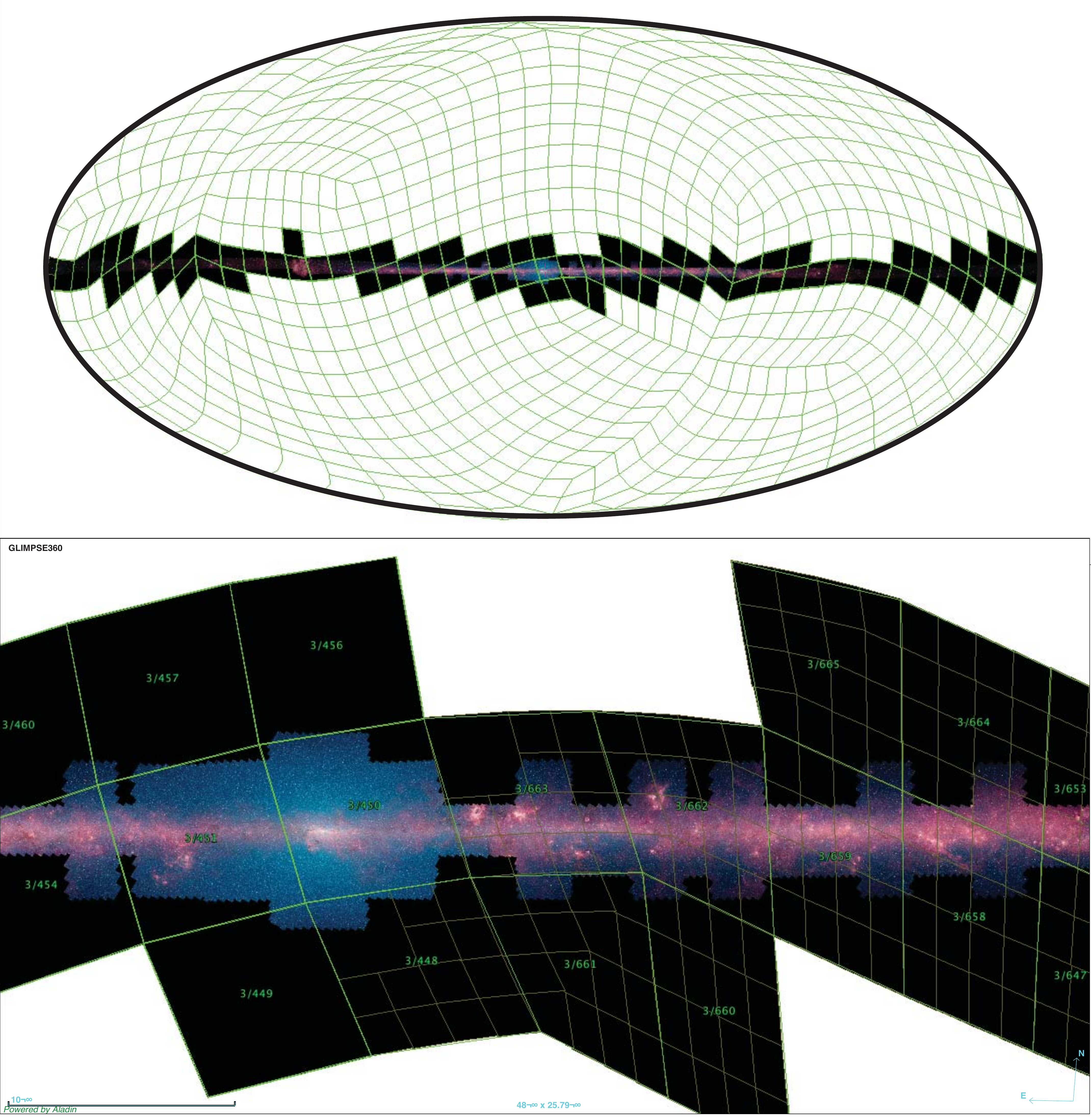}
\caption{HiPS of the Spitzer GLIMPSE 360 survey. The top panel shows a full sky Aitoff projection of the GLIMPSE 360 HiPS, rotated so that the galactic plane is horizontal and overlaid with a (green) grid of the lowest order $k_{tile}$=3 HiPS tiles with their tile index numbers. Note that only tiles that intersect the data are included in the HiPS and that regions of those tiles where there is no data are set to black. The lower panel shows a magnified view of the HiPS as shown by Aladin overlaid with the approximation of the HiPS tiles as they are projected onto the interactive display. The green grid shows how Aladin constructs the current view from $k_{tile}$=3 tiles, where in some parts of the view it is necessary to subdivide the tiles to gain more control points to minimise the distortion related to the bilinear tracing of the data onto the screen view.  }
\label{Glimpse_fig}
\end{figure*}

A visually stunning example is the HiPS that has been generated for the GLIMPSE 360\footnote{\url{http://www.spitzer.caltech.edu/glimpse360}} survey of the plane of the Galaxy. GLIMPSE 360 is a part of the Galactic Legacy Infrared Midplane Survey Extraordinaire (GLIMPSE, \citet{2003PASP..115..953B}, \citet{2009PASP..121..213C}) observed by the Spitzer mission  \citep{2004ApJS..154....1W}. The GLIMPSE 360 survey comprises images at 3.6 and 4.5 microns and the processed data are available from IRSA\footnote{\url{http://irsa.ipac.caltech.edu/data/SPITZER/GLIMPSE/}}.  A HiPS representation of the processed and mosaicked data has been generated\footnote{by Thomas Robitaille} using the CDS {\em Hipsgen} program. As the data were already stitched together as a high quality colour image, the construction of the HiPS concerned the mapping of the data onto the highest order grid of $k$=18, followed by computation of all of the orders down to $k$=12 with 512$\times$512 tiles (corresponding to a range of tile orders $k_{tile}$=9 to 3). The result shown in 
Figure~\ref{Glimpse_fig} illustrates how HiPS provides a whole-sky view of this data set, and how the tiles from different orders are used to visualise the data at different resolutions.

 For a number of the HiPS generated by the CDS there was a need to perform a background level adjustment of the original images. Simple subtraction of a constant background level has been applied to the 2MASS J-band survey using the original SKYVAL values.  In other cases the background level has been estimated from a subset of pixels of each original image. This has been done for the SuperCOSMOS H$\alpha$ survey for which no background level was provided with the original data sets. Other examples required the use of masks to remove bad (or non-data) pixels from the original images. The GALEX AIS survey images have a circular data area within the rectangular data arrays (because of the circular GALEX field of view), requiring the use of masks to select the data values to generate the HiPS.  

For data sets that are already in HEALPix format, the generation of their HiPS representations is straight forward as it only requires derivation of the set of maps for the desired tile orders from the original HEALPix map. This is the case for maps of the cosmic microwave background (CMB) radiation that are available via the Legacy Archive for Microwave Background Data Analysis (LAMBDA\footnote{\url{http://lambda.gsfc.nasa.gov}}).  

The WMAP Year 9 data, for example is provided by LAMBDA as a $k$=9 ($N_{side}$=512) HEALPix map with $\theta_{pix}$=6\farcm87. This map can be directly expressed as a single order HiPS with $k_{tile}$=3 and $n_{tile}$=64 by simply grouping the original HEALPix pixels into 64$\times$64 size tiles. Here there is no need for generation of lower orders because the data are already at the lowest useful HiPS tile order of three. 

The Planck LFI data have been provided by ESA/ESAC\footnote{European Space Agency / European Space Astronomy Centre} as $k$=10 ($N_{side}$=1024) HEALPix maps for the 20, 44, and 70 GHz bands, with the 70 GHz band also duplicated as a $k$=11 ($N_{side}$=2048) map. The HFI products at 100, 143, 217, 353, 545, and 857 GHz are provided as $k$=11 ($N_{side}$=2048) maps. As with WMAP, these Planck maps can be represented by single order HiPS with the $k$=10 LFI maps expressed as a $k_{tile}$=3 grid with $n_{tile}$=128. The $k$=11 LFI and HFI maps are naturally expressed as $k_{tile}$=3 grids with $n_{tile}$=256. See Table~\ref{table:2} for the properties of the HiPS tiles and pixels for HiPS with $n_{tile}$ of 64, 128, and 256.

Note that the original coordinate systems of all of the CMB data sets are Galactic coordinates. The HiPS of these data sets do not make any change in this respect as their HiPS representations simply use Galactic coordinate system HEALPix grids. 

The methods used to generate the HiPS described in this article are available in the CDS {\em Hipsgen} program. The preferred input files for {\em Hipsgen} are FITS images with valid WCS headers. Compressed (gzip, zip, hcompress) and multi-extension FITS files are also supported. The output formats can provide the complete dynamic range of the data or a more compact representation. The full pixel values can be provided using FITS format. Using JPEG or PNG output formats provide a more compact result that degrades the dynamic range of the pixel values to 256 grey levels (or 255 grey levels plus one transparency channel).

Colour images may be generated using two or three input data sets corresponding to RGB channels. It is possible to create coloured HiPS based on JPEG or PNG coloured tiles, either from an original coloured data set (for example the SDSS colour HiPS provided as a collection of JPEG images with an astrometric calibration), or from two or three already computed HiPS images where the individual tiles can be combined into colour RGB tiles.

The volume of data in a HiPS representation of a full sky survey is typically an increase of $\sim$1.5 times the volume of the original data. The total number of pixels over an entire HiPS structure will be limited to a factor of 4/3 ($\sum\nolimits_{n=0}^{\infty}\frac{1}{4^n}$) greater than the $N_{pix}$ of the highest order map because of the four-fold decrease in the number of pixels within each lower order. For example, summing $N_{pix}$ (from Table 1) over the $k$=18 to 12 orders used for the DSS HiPS shows that the total number of HEALPix pixels in this HiPS is increased by a factor of 1.33325 over the $k$=18 map. In practice the file storage of these pixels leads to a greater increase, for example the 1.472 TB of DSS original files becomes 1.39 times larger as a 2.048 TB HiPS. The file formats have a much more significant effect on the volume, with JPEG and PNG formats being 20-40 times smaller than the FITS representation (dependent on image depth and structure). 

\subsubsection{HiPS for collections of pointed observations}

The HiPS scheme is sufficiently general to handle data with any kind of sky coverage and there is no restriction on the contiguity of the input images. This facilitates the use of HiPS to combine collections of data, such as the pointed observations of a given instrument. The archives of the Hubble Space Telescope (HST) at the CADC have been used as a testing ground for the development of HiPS for pointed observations. Launched in 1990, the HST has acquired more than one million observations of which some 887000 are direct images.  Although HST is not a survey instrument, the quality and depth of most images are extremely valuable for a vast variety of astronomical projects, and we consider that the HiPS system for HST will open up new ways of using these images for research.

To compute the HiPS for public HST images one has first to decide how those images should be grouped together. Consideration of the distribution of the observations in the many different filters (and filter combinations) shows that HiPS generated for a subset of $\sim$15 filter groups would cover most of the HST observed fields. In defining these groups it makes sense to combine similar bandpass filters into a single group based on the number of images that use those filters and their characteristics. In addition, it is intended to eventually provide an achromatic HiPS for all the HST images. For each of the selected 15 filters groups we extracted the calibrated science data as input for the HiPS (using the CADC HST cache system).  This was done for single images and for the already combined composite images (associations).

Extensive trials were done to find the best parameters to generate this first set of HST HiPS. The biggest challenges were to find the best choices to manage the sky brightness, and to handle the small fields of view.  For sky brightness, we find that using an automatically estimated background value for each image produces a more uniform visual product. The HST instruments ACS, WFC3, NICMOS, and WFP2 have intrinsic angular resolutions between 50 mas and 0\farcs1, so we tested different values of $k_{max}$ for the highest order maps of these data. Following these tests we chose to over-sample the data by using a maximum pixel order of $k_{max}$=23, which has $\theta_{pix}$=25.1 mas (with corresponding tiles of order $k_{tile}$=14, $\theta_{tile, 512}$=12\farcs1). Presently we use a single set of parameters to generate all of the HST HiPS, and this is done with the {\it Hipsgen} program\footnote{
java -Xmx6000m -jar AladinBeta.jar -skyval=true -blank=0 -force -in=srcDir -out=trgDir -order=14 -img="reference\_image" -fitskeys="INSTRUME OPT\_ELE DETECTOR TIME\_EXP PROP\_ID PRODTYPE TARGNAME NMEMBERS OBS\_DATE"}.

During the trials we found that it was extremely useful to preserve the access links to the original images that contribute to a given HiPS tile, along with a small set of metadata that describes the images. Given the size of these data collections, it is important for the practical (data access) efficiency to extract this information while computing the HiPS. This metadata includes the exposure time, proposal number, observation date, and other information which is then stored with the tiles. This information can then easily be used alongside the tile images when visualising the HiPS, such as to provide a link (e.g. as an overlay) on each tile to the original data in the archive. 

Generating HiPS from large data collections such as HST requires significant planning and consideration of the growth and maintenance of the system. As the HST is still in operation, the CADC has decided to re-generate the HiPS for each of the filter groups every six months or so. Systems for incremental updates will likely be needed to streamline such operations.

\subsubsection{HiPS images and MOC}

MOC maps may be used to support HiPS data sets in a number of ways. A MOC map that has been constructed from a HiPS data set at the resolution of the smallest HiPS tile provides a convenient summary of the coverage of a HiPS data set. The construction of a MOC at any chosen HiPS tile resolution is straight forward because a HiPS map contains only the tiles that have data values, and the list of tiles may be used as the initial list of HEALPix pixels for the definition of its MOC. As described in Section~3, the initial list is then compacted to build the MOC.  In this way, MOC maps of any chosen resolution may be constructed. 

We found that it was useful to generate a low-resolution MOC maps for each of the HiPS data sets hosted by the CDS, where the resolution of each MOC is chosen to be about four orders lower than the highest order (i.e. k$_{max}$-4) of the HiPS. These MOC maps are stored with the metadata of the HiPS. The ease of generating such MOC maps has facilitated the implementation of interactive MOC generation tools in Aladin, which includes the capability to specify and change the MOC resolution. 

\subsection{Catalogues}

\begin{figure}
\includegraphics[width=8cm]{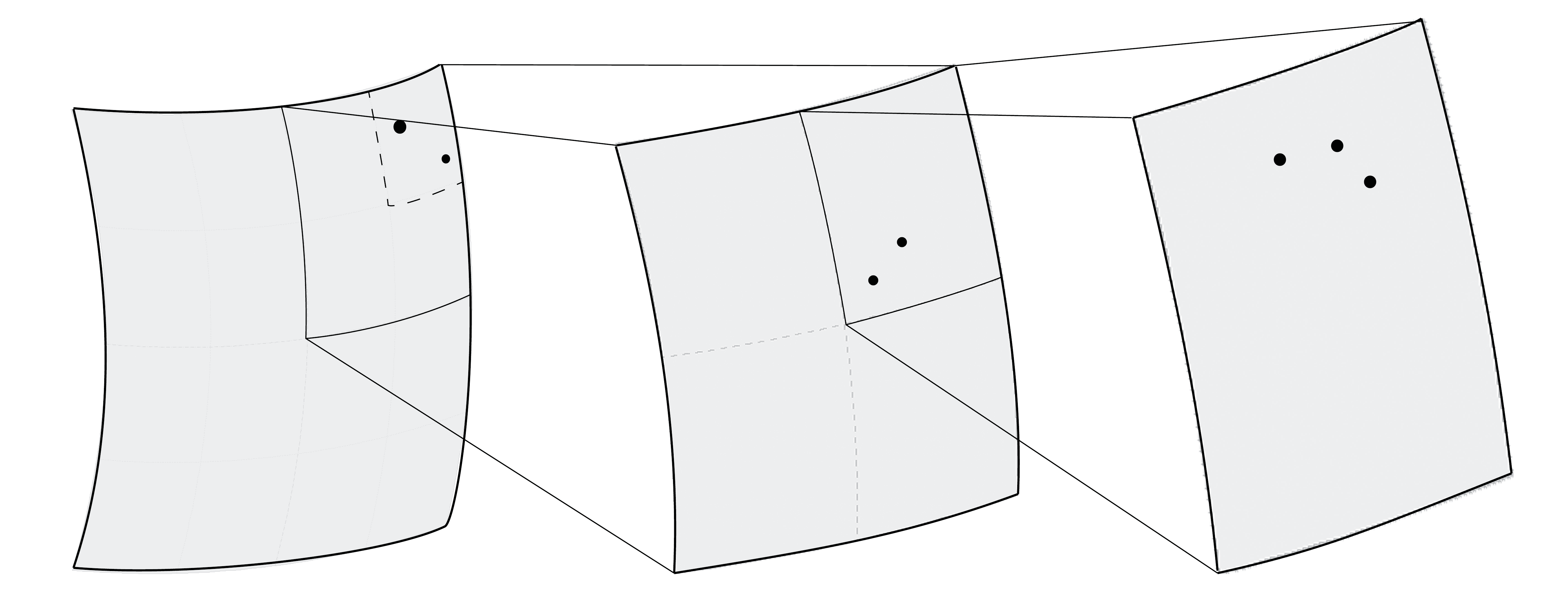}\\
\includegraphics[width=8cm]{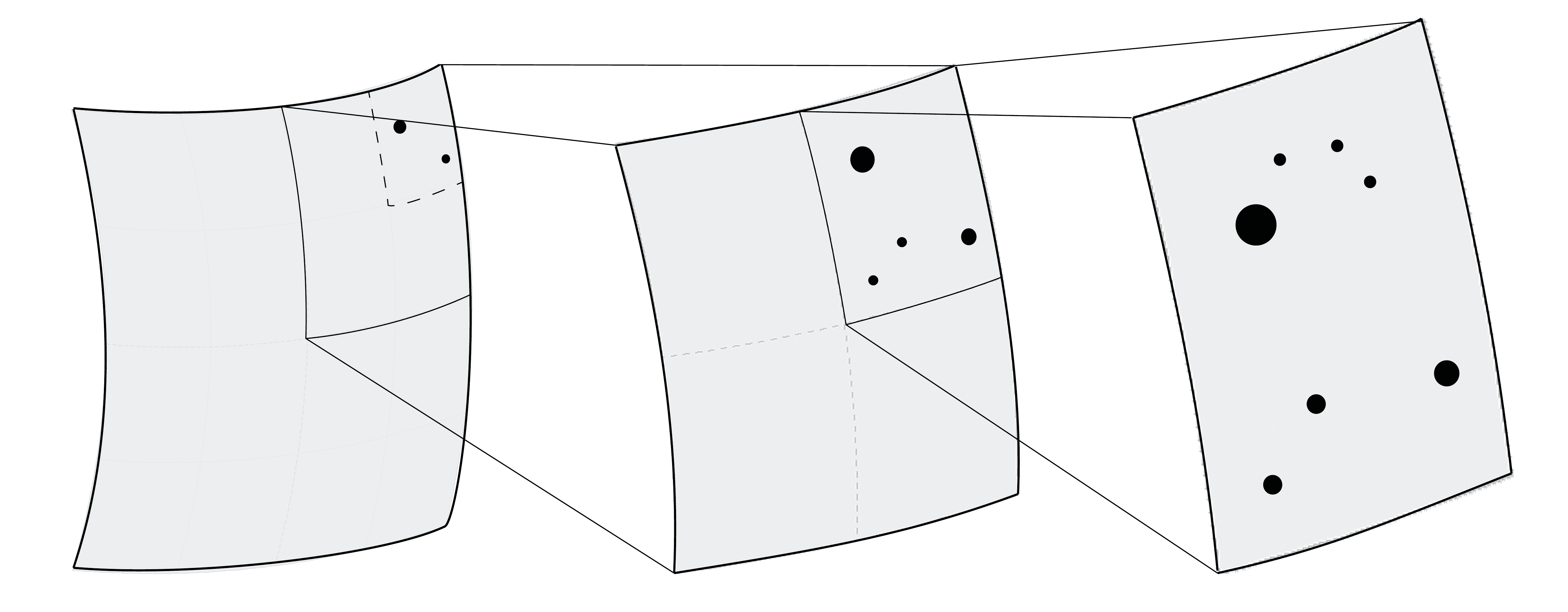}
\caption{HiPS catalogue tiles -- content and visualisation. The content of a HiPS catalogue tile depends on how the source catalogue points are chosen to be distributed over the different HiPS orders. The upper panel shows three HiPS tiles of successive orders, with the lowest order tile on the left and sibling tiles to the right. The filled circles in the tiles represent catalogue sources that are 'stored' in each tile, and the upper set of tiles represents a {\em sequential view} of these orders. The lower panel shows the same three tiles but in this case we represent a {\em cumulative view} of the HiPS catalogue, where zooming into higher order tiles shows the sources in the tile plus the cumulative set of sources over all of the lower orders. }
\label{Cat_fig}
\end{figure}

The development of HiPS for catalogues is aimed at enabling the manipulation and visualisation of large ($\sim$10$^{9-12}$ source) catalogues. HiPS provides a means to organise a catalogue based on the spatial distribution of the source catalogue points on the sky. A HiPS catalogue made up of multiple tiles over a range of tile orders ($k_{tile}$) can provide different views of the catalogue that change as a function of the angular resolution. The freedom to choose which subset of the catalogue to include at any given order means that, in addition to the spatial organisation, the catalogue may also be organised over the different orders according to a hierarchy that can be based on any quantitative property of the catalogue, such as the source brightness, or field density, source redshift, or distance. The association of subsets of catalogue sources with HiPS tiles over a range of orders enables a \emph{progressive} view of the catalogue so that the selection of sources that are displayed can change as a function of the zoom level. For example, a HiPS catalogue organised by source brightness can be presented so that the widest full sky view shows only the brightest sources, with fainter sources progressively appearing in the display as one zooms into smaller and smaller regions (Figure \ref{Cat_fig}).  

In practice a HiPS catalogue is organised in the same way as a HiPS image survey, making use of the HiPS tiles to organise the spatial sky coverage of the catalogue so that the astronomical source positions in the catalogue are associated with the corresponding HiPS tiles. Whereas each tile of a HiPS image survey contains a 512$\times$512 image, each tile of a HiPS catalogue contains a list of catalogue sources (with at least the RA and declination coordinates of the sources) in the form of catalogue rows that have been extracted from the full original catalogue. In the present implementation of HiPS catalogues the tiles are stored as ASCII tab separated value (TSV) files, and these tiles are organised into the file system directory structure of HiPS tile orders. The metadata about the columns of the catalogue are stored in a VOTable \citep{2011arXiv1110.0524O} document alongside the lowest order tile files.

The construction of the catalogue tiles involves extracting rows from the original catalogue into the HiPS tiles over a defined range of orders. The main considerations when constructing the catalogue tiles are to decide on the approximate number of sources per tile, and the definition of the sorting key. The sorting key determines the distribution of the catalogue sources over the HiPS tiles as a function of the HiPS order, and the sorting key can be defined by any quantitative property. 

Given the sorting key, catalogue tiles can be constructed by successively filling the tiles with sorted order sources up to an approximate limit for each tile, and then recursively moving to the next order to fill the four sub-tiles.  Practical tests show that catalogue source densities of up to $\sim$500 sources per tile allow for manageable zooming, but the tile source limit need not be constant, and in some cases it has been found to be useful to allow logarithmic or other non-linear increases in the number of sources per tile as a function of the order.  We note that HEALPix aliasing patterns can occur when the original catalogue has heterogeneous distribution, and that this can be minimised by modifying the number of sources in the tiles based on the catalogue source density on wider scales.

Filling the tiles in this manner means that each catalogue source is only included once in the HiPS catalogue. Visualising a HiPS catalogue by scanning through the orders provides a progressive view of the catalogue, and this can be controlled in a number of ways. For example, a sequential view of the tiles of different orders would show only the sources in each order. More commonly,  a cumulative view shows all of the sources in the area of the tile up to a given order. This is illustrated symbolically in Figure~\ref{Cat_fig} where the top panel represents a sequential view of three successive orders of a HiPS catalogue, showing the catalogue sources that are stored in each order. The lower panel shows a cumulative view of the same HiPS catalogue, where zooming into successive higher orders shows all of the sources in the area of the tile up to the current order. Note that construction of a cumulative view requires combining all of the sources from the current tile and all of its parent tiles.

The 2MASS all-sky catalogue of point sources \citep{2003tmc..book.....C} contains $\sim$470 million sources with measurements in the  J, H, and Ks bands.  A HiPS representation of this catalogue has been generated by mapping the positions of the sources onto HiPS tiles up to a maximum order of $k_{tile}$=11, which corresponds to a tile size of 1\farcm72. The sorting key for this HiPS catalogue is based on the infrared brightness of the sources where we have combined the fluxes in the J, H, and Ks bands.  

The largest HiPS catalogue that has so far been constructed is that of the Gaia universe model snapshot \citep{2012A&A...543A.100R}.  This is based on GUMS-10 simulation of the expected content of the catalogue that will be generated from the ESA Gaia astrometric mission \citep{2001A&A...369..339P}. Gaia is expected to create a catalogue of $\sim$10$^{9}$ stars with astrometric accuracies of 5-15\ $\mu$as (for G band magnitudes $<$12) and 5-600 $\mu$as for the full catalogue. The $\sim$2$\times10^{9}$ simulated Galactic objects in GUMS-10 catalogue of "milky way stars" are based on the Besan\c{c}on model \citep{2003A&A...409..523R}, which provides the distribution of the stars, their intrinsic parameters, and their motions. The HiPS of this catalogue uses the simulated sky coordinates and the sources are distributed over HiPS orders up to k$_{tile}$=11. The sorting key is chosen to be the simulated barycentric distance (in parsecs) that is provided in the catalogue, a quantity that the mission is expected to derive from accurate parallax measurements. 

Another example is the HiPS catalogue that has been constructed from the CDS SIMBAD\footnote{\url{http://simbad.unistra.fr}}
astronomical database \citep{2000A&AS..143....9W}. SIMBAD provides basic data, cross-identifications, bibliography, and measurements for astronomical objects. SIMBAD currently includes over 7.6 million well documented astronomical objects and the database can be queried by object name, coordinates, and various criteria. While SIMBAD is not a single catalogue, much of the information can be expressed in the form of a catalogue of sources so that SIMBAD may be represented by a HiPS catalogue. Experimenting with different sorting keys for distributing the SIMBAD sources over a range of HiPS orders showed that sorting by properties such as source brightness, spectral properties, angular size, or literature statistics all provide new and useful ways of viewing SIMBAD.  The publicly available SIMBAD HiPS catalogue is sorted by the number of literature references associated with each astronomical object. The resulting HiPS employs tile orders up to $k_{tile}$=14, which has a tile size of 12\farcs9. When viewed in Aladin the SIMBAD HiPS shows the most highly referenced \emph{famous} sources in the zoomed-out wide scale view, with more sources appearing as the display is zoomed into higher orders.

The methods used for generating the HiPS catalogues described above are all included in the CDS {\em Hipsgen-cat} program. {\em Hipsgen-cat} accepts inputs in the form of list of sources in CSV, FITS, or VOTable format, and has been successfully tested with catalogues of up to 2 billion sources. The typical generation time is a few hours for catalogues of the order a hundred million rows, for a maximum HiPS tile order of $\sim$11.

\subsection{Cubes}

\begin{figure}
\includegraphics[width=8cm]{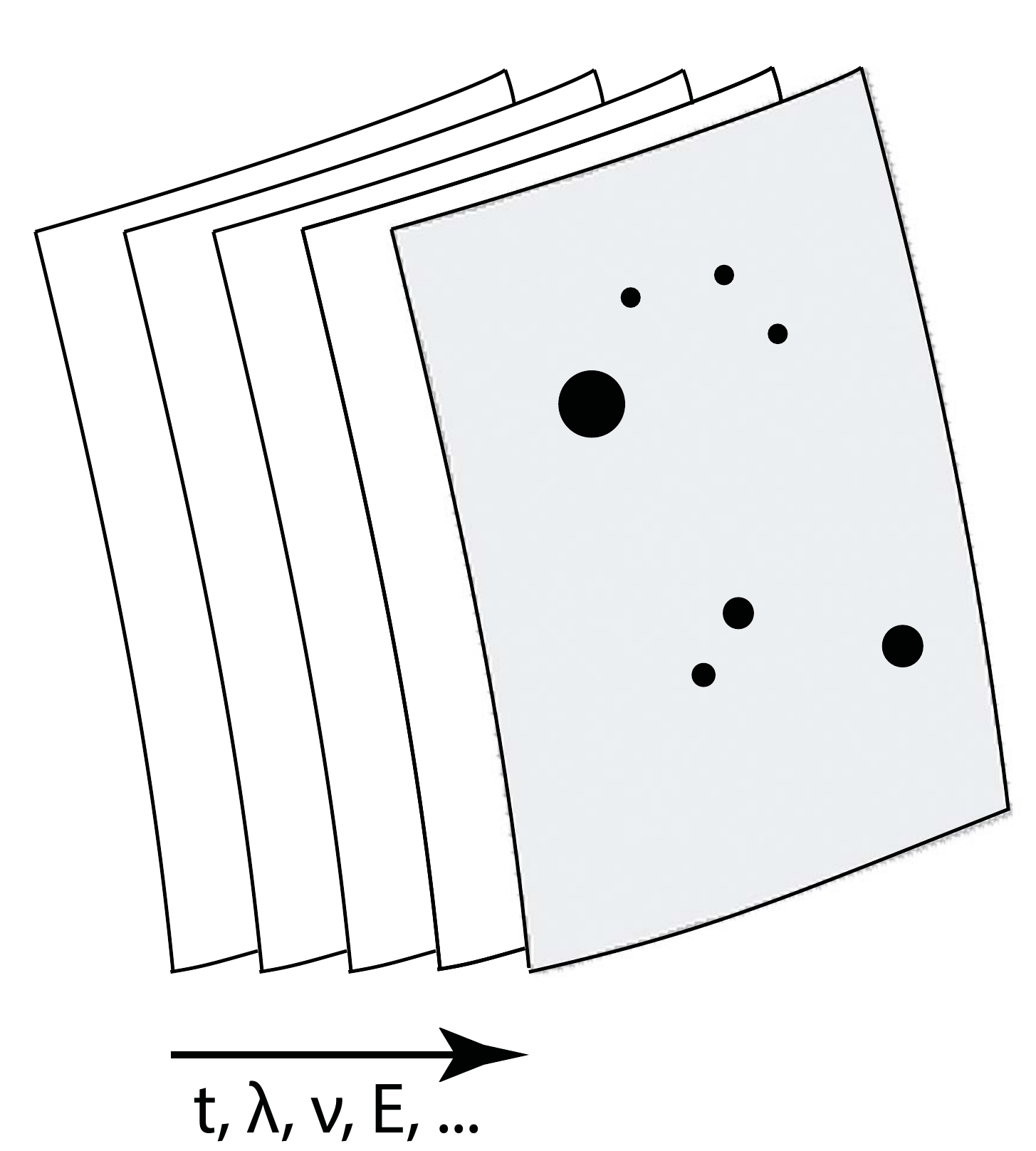}
\caption{HiPS cubes. HiPS cubes are simply a set of nested HiPS images, where the third dimension of the cube may represent time ($t$), wavelength ($\lambda$), frequency ($\nu$), Energy ($E$), or any other quantitative parameter. HiPS stores the third dimension planes in the same way as HiPS tiles, using simple files, where the successive planes are labelled with a sequential suffix number. }
\label{Cube_fig}
\end{figure}

The development of HiPS cubes extends the concept of HiPS images into a third dimension to enable the hierarchical organisation of three-dimensional data cubes which have two spherical coordinate axes plus a spectral \emph{(wavelength, frequency, energy)}, velocity, or temporal axis as is symbolically illustrated in Figure~\ref{Cube_fig}. The basic idea here is that each channel (or plane) of the third dimension of the data cube is stored in additional HiPS tiles.

The hierarchical description of the spatial part of HiPS cubes is identical to that of HiPS images. Experimentation with different axis compression schemes for the third dimension using a variety of different cube data showed that it is preferable to retain all of the third dimension planes, even when the spatial dimensions are heavily smoothed as occurs for low order HiPS tiles. This choice is mostly guided by tests of spectral cubes, where we found that visualisation of the sharp spectral features in the cubes is important even when the spatial dimensions were compressed, and that the blurring of spectral features in compressed data negated the benefits of the zoomed-out views of the cube. 

The number of third dimension planes in astronomical data cubes can vary greatly; cubes constructed from multi-band images may have only a few planes, whereas radio spectral data cubes and integral field spectroscopy cubes may have hundreds or thousands of planes.  To avoid streaming of large volumes of data from HiPS cubes, as might occur for the low orders of HiPS cubes with many planes,  we find that sampling of the third dimension to access only every $n$th (2nd, 5th etc.) plane is a satisfactory solution for low HiPS order projections. As such the control over the effective reduction in the number of planes to be downloaded is given to the client programs.

This approach also effectively prioritises fast spatial manipulation of the data ahead of spectral operations that might be performed by orthogonal cuts or extractions along the third dimension. By unpacking the cubes into separate frames we enable fast access to the individual planes of the cube. We have tested an alternative storage of the cubes in more integrated cube formats such as MPEG, which would allow for faster access (and compression) along the third dimension, but we found that the extraction of individual frames from multiple MPEG formatted cubes of adjoining tiles was too slow for the spatial manipulation of the cube.

The construction of HiPS cubes is similar to that of HiPS images, with the extra dimension taken into account via the use of multiple frames for each tile image. These frames are simply multiple tile images in separate files and enumerated using a sequential suffix number, 0 to N(cube planes)-1. As is the case for HiPS images, the cube frame tiles may be stored as FITS, JPEG, or PNG files. For data that is already in cube format we unpack the planes to produce HiPS tiles for each plane. For multi-band surveys HiPS generation is straight forward, but care must of course be taken with the astrometric correspondence of all the band images. As such, HiPS of multi-band surveys may be simply constructed a posteriori from the HiPS of the individual bands.

A HiPS cube has been generated for the Canadian Galactic plane survey (CGPS, \cite{2003AJ....125.3145T}). The CGPS is a wide area (1600 square degrees) atomic hydrogen survey covering a large fraction of the Galactic plane with an angular resolution as small as 1\arcmin. The original observations comprise 373 fields with the Dominion Radio Astrophysical Observatory (DRAO) synthesis telescope, and the processed calibrated data\footnote{available from the CADC at \url{http://www3.cadc-ccda.hia-iha.nrc-cnrc.gc.ca/en/cgps}} consist of sixty-eight 5\degr $\times$ 5\degr (1024$\times$1024 pixel) data cubes with 272 spectral channels with velocity resolution of 1.3 km\ s$^{-1}$.  The HiPS cube of this data was generated by mapping all of the image planes of the data cubes onto 272 HiPS structures using tiles of size 512$\times$512, with a range of orders up to $k_{max}$=14 corresponding to a maximum resolution of 12\farcs9. The highest order tiles are  $k_{tile}$=5 which have an approximate size of $\theta_{tile,512}$=1\fdg 83. The 272 HiPS structures are stored in an integrated way by combining all the files into a single HiPS structure and enumerating the tiles (with suffixes 0-271). The result is a homogeneous wide area HiPS cube that can be displayed \emph{all at once} in a tool such as Aladin. The HiPS contains precomputed image tiles of all of the CGPS cube planes over the range of orders, so in addition to the wide scale view, we may zoom into the cube to access the higher resolution tiles, and at any order we may also interactively scan through all 272 spectral channels. In this way HiPS enables the CGPS data set to be accessed and visualised more easily than by loading the sixty-eight fixed-resolution individual data cubes.

HiPS cubes have been constructed for the Calar Alto legacy integral field area (CALIFA) survey \citep{2012A&A...538A...8S}. The CALIFA survey is an on-going integral field spectroscopy survey of 600 local (0.005$<z<$0.03) galaxies obtained with the PMAS/PPak integral field spectrograph mounted on the 3.5 m telescope at the Calar Alto observatory. A HiPS cube was built from the 100 low spectral resolution (V500, R$\sim$850) cubes of the first data release \citep[DR1,][] {2013A&A...549A..87H}, and another HiPS cube was built from the extended (and superseding) sample of 200 galaxies in the second data release \citep[DR2,][] {GarciaBenito:2014wg}. The CALIFA observations have 74\arcsec$\times$64\arcsec\ hexagonal fields-of-view with 331 science fibres that project 2\farcs7 diameters on the sky with gaps between the fibres that result in a filling factor of 0.6. The processed data cubes have 1\arcsec $\times$1\arcsec angular sampling, and a linearly sampled wavelength axis with 1877 channels that span the wavelength range 3749-7501$\AA$. The HiPS cubes are constructed from the processed data, taking advantage of the consistent format and cube dimensions to generate HiPS images for each of the 1877 channels, and combining these into a single HiPS structure. Tiles of 512$\times$512 pixels are used, and the maximum order of the tiles was chosen to be $k_{tile}$=9 corresponding to a resolution of 0\farcs81 to slightly over-sample the processed data. In the few cases where multiple CALIFA cubes overlap on the sky the average flux values are used.  

The CALIFA cubes cover only a small fraction of the sky, but their combination into single whole-sky HiPS structures (one for each of DR1 and DR2) brings about a number of benefits. HiPS provides a convenient way to access all of the CALIFA data with one action to reveal the distribution of CALIFA observations across the sky. The supporting metadata, including the sky coordinates of the original data cubes, provides a way to easily select and zoom into the sky regions covered by the data. Zooming into the higher orders of the HiPS cubes shows the details of the data including the spectral channels. Furthermore, multiple views of the same HiPS structure allows direct comparison of the different CALIFA galaxies, and allows one to interactively scan through the spectral channels of multiple cubes at the same time. In addition to the positions, the metadata also includes the links to the original data so that the full individual cubes may be loaded alongside their HiPS representations. Also the MOC that is included in the metadata can be used to make general queries for other data that intersects the same region of the sky.

HiPS cubes have been constructed for a number of multi-band image surveys, where the planes of the cubes are the wavelength ordered bands of the survey.  A good example is the HiPS cube of the IRIS (improved reprocessing of the IRAS survey) data \citep{2005ApJS..157..302M}. To construct this HiPS cube the HiPS images of the individual 12, 25, 60, and 100 $\mu$m bands were combined into a single HiPS cube structure.

\section{HiPS data sets and tools}
\label{HiPS_avail_section}

\subsection{Currently available HiPS data sets}

There are presently $\sim$200 publicly available HiPS data sets. Lists of the current HiPS images, HiPS catalogues, and HiPS cubes are provided in Tables 3, 4, and 5 respectively. 

Table~3 represents $\sim$180 HiPS image data sets. These are organised in the table by waveband: gamma-ray; X-ray; ultraviolet; optical; infrared; radio bands; and categories for emission line surveys of H$\alpha$, \ion{H}{I}, and CO. Each HiPS has a unique CDS identifier. The table shows the base names of these identifiers, and each row represents either a single HiPS or a group of related HiPS. For example, the first row describes the four HiPS images of the Fermi data, one for each of the three energy ranges (3-300 MeV, 300-1000 MeV, and 1-3 GeV) and another one for the colour composite HiPS of this data. 

Table~3 lists the properties of each HiPS: the maximum tile order $k_{tile}$; the size of the smallest pixel $\theta_{pix}$; and the coverage of the HiPS as a percentage of the whole sky. Note that each HiPS image structure contains all of the tile orders from the maximum order down to $k_{tile}$=3, so for example, each of the XMM HiPS with maximum values of $k_{tile}$=7 are comprised of 5 successive tile orders $k_{tile}$=3,4,5,6, and 7. The deepest HiPS in terms of the number of orders is that of the HST archive images, which has a maximum tile order of 14 and corresponding pixel size of $\theta_{pix}$=25.1 mas.  

HiPS for which the maximum tile order is  $k_{tile}$=3 are single order HiPS of relatively low angular resolution ($\sim$1\arcmin-7\arcmin) such as the gamma-ray surveys, early X-ray surveys, CMB maps, and some radio astronomy surveys. There are 71 such single order HiPS (within 19 groups) in Table~3.

There are 77 full-sky HiPS in Table~3. The highest order HiPS that cover the whole sky are the 2MASS and DSS HiPS with maximum tile order of $k_{tile}$=9 ($N_{side}$=2$^{18}$) with $\theta_{pix}$=0\farcs81. There are 55 HiPS that cover less than 5\% of the sky, and 7 HiPS with less than 1\% sky coverage.

Table~4 lists the HiPS catalogues that are currently available. All of these have been generated from catalogues and data from the CDS Vizier and SIMBAD services. For each of these data we tabulate the CDS identifier, a brief description, and references to the original published catalogues. 

Table~5 lists the data sets that have been converted into HiPS cubes. These represent a wide range of cube characteristics in terms of the number of planes, field of view size, and distribution over the sky. 

Most of the HiPS in Tables~3, 4, and 5 have been created by the CDS from original or processed data that have been obtained from publicly available archives, or via collaborations with individual authors, projects, or partner data centres. There are 35 HiPS that are hosted outside the CDS, and there are 20 HiPS that have been computed externally to the CDS but that are hosted by the CDS.  The table notes specify the sources of the original data and identify the parties who created the HiPS representations. We also also indicate the data centres where each HiPS is hosted. References to the published articles on the surveys, instruments, data releases, data processing, and data characteristics are also provided where possible.  

The lists of the currently known HiPS images, catalogues, and cubes are maintained at the CDS in the HiPS directory\footnote{\url{http://aladin.unistra.fr/hips/list}}. The HiPS directory provides information on each of the HiPS similar to what is presented in Tables 3, 4, and 5, along with other practical information including the full CDS identifier and the root URL of each HiPS. The file formats that are available for each of the HiPS are also indicated, along with the date of last modification. The HiPS directory thus provides an overview of the current status of all HiPS data sets.

\subsection{Tools for accessing and visualising HiPS}
\label{Tools_section}

There are currently three astronomical visualisation tools able to visualise HiPS images. These are the Aladin desktop Java application and the Aladin Lite Javascript web widget both developed by the CDS, and Mizar which is a WebGL client developed by CNES as part of their SiTools 2 publication framework \citep{2011ascl.soft11008M}.  

To visualise HiPS images a tool needs to perform a number of steps. Starting from the minimal information of the HiPS root URL, a tool needs to read the properties file of the HiPS to retrieve metadata about the maximum order, the size of the tiles, and the coordinate frame (ICRS or Galactic). To display the HiPS for a region of the sky the tool must compute the tile order that is best suited to the desired field of view. Then, using the HEALPix library \citep{2011ascl.soft07018G}, and knowledge of HiPS tiling, the tool needs to identify which tiles it needs for the view,  and to download the files for those tiles. 
To draw the tiles on the display, the display coordinates of the four corners of each tile must be determined, and then the tile image can be distorted through affine transformations to map it onto the display area defined by the four mapped points.

Aladin is a multi-featured Java application for the discovery, access, and visualisation of astronomical data with detailed functionality for images and catalogues. Aladin is VO compatible, and can serve as a portal to access data made available by VO protocols. The early development of Aladin is described in \citet{2000A&AS..143...33B} and current information is maintained at the Aladin web pages\footnote{\url{http://aladin.unistra.fr}}. The HiPS visualisation capabilities of Aladin have been developed alongside the development of the HiPS system at the CDS, and Aladin is the primary tool for testing and verifying HiPS.  Aladin provides an interface to access all of the HiPS described in the HiPS directory, and it can also be used to access any available  HiPS given its local file address or URL address. Aladin provides a familiar all-sky zooming and panning interface that makes full use of the multi-resolution features of HiPS. Multiple HiPS may be viewed simultaneously in separate windows, or layered together with transparency overlays, and views may be locked together for synchronised zooming and panning. For HiPS cubes, Aladin provides a facility for stepping through the third dimension of the cube plane by plane or as an animation. HiPS catalogues may be overlaid on images and zoomed to view the different HiPS catalogue orders. The general Aladin catalogue overlay feature also provides a very useful mechanism for going from a view of the HiPS image, to the original data sets that are listed in the \emph{progenitor} metadata catalogues that can be included in HiPS. Aladin also provides a set of functions for using MOC maps (e.g. Figure~3 shows screen shots of Aladin being used to overlay a MOC map on the HiPS of the SCUBA data) including a multi-resolution display and the ability to combine MOC maps with logical operators to form unions and intersections. 

In addition to its capabilities for \emph{using} HiPS and MOC, Aladin also includes all of the tools for \emph{creating} HiPS and MOC maps from images, catalogues, and cube data sets. All of the HiPS described in this paper have been generated with the {\em Hipsgen} tool that is provided in the standard Aladin desktop application. {\em Hipsgen} can be used from the command-line or directly from the Aladin tools menu. 

Aladin Lite is a simplified HiPS visualiser that is developed in the JavaScript programming language. It runs in a web browser (without the need for a plug-in) and is able to display HiPS images (in JPEG or PNG formats), and HiPS catalogues with zooming and panning capabilities similar to the Aladin desktop application. Other tabular data and information can also be overlaid.  Aladin Lite can be used to visualise local HiPS data and HiPS that are hosted on any web server including those in the HiPS Directory. A strength of Aladin Lite is that it can be embedded easily on a third-party web page \citep{2014ASPC..485..277B} and it comes with an application program interface (API) that provides the necessary controls to customise Aladin Lite to different needs. A good example of this is the Spitzer web page\footnote{\url{http://www.spitzer.caltech.edu/glimpse360/aladin}} of the GLIMPSE360 project where Aladin Lite is used (alongside WWT) to visualise a colour HiPS of the GLIMPSE360 survey data.  Other examples include the Aladin Lite implementations at CADE\footnote{\url{http://cade.irap.omp.eu}} to preview HEALPix maps of various surveys, and the ADS\footnote{SAO/NASA Astrophysics Data System} all sky survey\footnote{\url{http://www.adsass.org}}.

\section{Discussion}
\label{Discussion_section}

Concepts for the hierarchical organisation of astronomical data are an important part of how astronomy approaches the challenges of big data. The main ideas are that i.) hierarchical systems describe the data in a multi-resolution manner that facilitates efficient access to the data, ii.) hierarchical systems scale easily so that they may handle large and small data sets in the same framework, and iii.) the use of a generic hierarchical system for the wide variety of astronomy data enables new levels of interoperability. 

HiPS has been built with these underlying concepts in mind, and the result is a practical multi-resolution HEALPix data structure for astronomical images, catalogues, and three-dimensional data cubes. Here we discuss various visualisation and scientific aspects of HiPS. An important distinction of HiPS is that it goes beyond visualisation to take the scientific considerations of astronomical data into account providing a new statistical view of the data.  Also, HiPS is designed to be easy to implement and we outline how it can be used to enable data sharing and interoperability.  

\subsection{Scientific visualisation} 

The description of HiPS in this paper has been largely presented from the point of view of visualising astronomical data. This is a convenient way to describe HiPS and indeed a number of the characteristics of HiPS, such as the preferred tile size, are motivated by practical considerations related to accessing and streaming of multi-resolution astronomy data to a visualisation tool.  In terms of supporting visualisation of astronomical images, HiPS has similar performance to the other hierarchical formats that are used for this purpose (TOAST/HTM). The HiPS system based on HEALPix, provides a means to conserve the scientific properties of astronomy data, and goes beyond the description of images to extend the hierarchical description of data to catalogues and three-dimensional data cubes so that these data types may also be visualised as multi-resolution structures.  

\subsection{HiPS facilitates scientific use of astronomical data}

To make HiPS a hierarchical framework that supports scientific use of data we considered that it was important to i.) use a sky tessellation that was directly translatable into astronomical coordinates, ii.) support accurate photometric properties of images that would allow scientific computation directly on the hierarchical grid pixels, and iii.) provide a way to link directly to the original \emph{progenitor} pixel or catalogue data.

HEALPix has proved to be a good choice because it automatically supports the first two considerations, and the catalogue and metadata support mechanisms of HiPS provide a simple way to include links to the original data. Also, HiPS benefits from the scientific computing libraries and tools that have already been developed for HEALPix. These libraries  and tools are largely from the CMB and extended-source astronomy community, but there is also increasing use of HEALPix for other projects, for example Gaia is using HEALPix \citep[private communication William O'Mullane;][]{MignardDPAC}. There is much potential for the wide spread use of HiPS and integration of HiPS functions in astronomy software. This would build on our earlier work to define MOC \citep{Boch:2014vv} that has resulted in MOC functions being included in the official HEALPix library \citep[][private communication Martin Reinecke]{2011ascl.soft07018G}. 

HEALPix, and hence HiPS, provides a direct translation from the pixels to the astronomical coordinates. HiPS supports both Galactic and equatorial (ICRS) coordinate systems. This duality is in response to the fact that many data sets have a natural coordinate system of Galactic or equatorial, leading us to support both systems. All of the current HiPS visualisation tools support both systems so that Galactic and equatorial HiPS can be used together. Computational interoperability between HiPS however requires the conversion of the data to one system or the other, and there may be cases where it will be necessary to host both Galactic and equatorial versions of HiPS.

HiPS manages both visualisation and scientific computational considerations by allowing the HiPS tiles to be encoded in both visualisation and science data formats. HiPS tiles can be duplicated in multiple formats, and these are simply stored alongside each other in the HiPS file structure. FITS files are typically used for science grade data with the full original dynamic range of the data values encoded in FITS in the standard manner. The efficient transfer and display features of the JPEG and PNG formats facilitate their use in visualisation systems. Having both means that one can visualise the data efficiently, and then seamlessly switch to the strict science data format to access the actual measurement values and perform computations. Other formats or compression schemes \citep[e.g.][]{McEwen:2011iz} could be explored as alternative tile encodings.

Another way in which HiPS facilitates the interaction between data visualisation and analysis is by providing a mechanism for direct links to the progenitor original data. The information about the original data is best included into the HiPS at the time of generation (especially for large data sets). This facilitates making catalogue style overlays of clickable links to the original archived data. This is also supported by tools that can display the original FITS data alongside of the HiPS representation.

In terms of the photometric accuracy of HiPS, the successive orders generated by simple averaging (as usually done for FITS format HiPS) are strictly flux conserving. The initial construction of the highest order map is of course a re-sampling of the original data (except for data that is originally in HEALPix format), and the photometric characteristics of the data are determined by that re-sampling. While the tools provided in Aladin for generating HiPS ({\em Hipsgen}) provide basic resampling and mosaicking capabilities, there can be advantages to pre-mosaicking data with more control over the photometric properties using tools such as SWarp \citep{2002ASPC..281..228B} or Montage \citep{2010ascl.soft10036J}. We also note that \citet{2012A&A...543A.103P} have already made progress with drizzle techniques for HEALPix data. 

\subsection{Interoperability of images, cubes, catalogues, and coverage maps}

HiPS and MOC allow a new level of interoperability between image surveys, data cubes, catalogues, and sky coverage maps. The common basis of HEALPix means that both large and small scale data sets may be easily visualised and compared. The expression of sky regions as MOC maps facilitates the comparison of sky regions covered by different data sets (of any type) such as to establish regions of intersection between multiple surveys. With MOC, such comparisons reduce to the simple comparison of lists of integer HiPS tile numbers, and this method scales very well when there are many data sets to be compared. MOC is also very useful for establishing whether a given sky coordinate falls within the coverage of a survey, or within the instrument fields of view of observations in an archive. These capabilities also extend to the computation of the intersections between survey areas and sources in large catalogues. 

Furthermore, services that provide access to data with a given sky coverage may employ a MOC description to vet in-coming (coordinate based) queries. A service may make an initial comparison of the coordinates of an in-coming query with the MOC map coverage of the data.  This allows the service to provide a fast initial response that would indicate whether a detailed database query of the service is likely to yield a result, hence avoiding complex queries on services that would return null results because of un-matched sky coverage.

\subsection{Implementation of HiPS}

The use of a simple file and file system structure to organise the tiles of a HiPS makes it very easy to publish HiPS to the web. The simplicity of such systems is a very practical consideration as it significantly lowers the barrier to implementing HiPS compared to systems where special databases or web servers are required. The use of simple files also means that the system is transportable and scalable, and can be easily duplicated on multiple servers (for example as mirror copies). Such transportability facilitates the local generation and testing of HiPS before deployment on public systems. HiPS is scalable in that the individual component files are small (i.e. managing large HiPS data sets involves managing large sets of small files) so that increases in volume are not affected by file size limitations. 

With the growing number of HiPS currently being generated by different data centres, projects, and individual astronomers, it is important to be able to organise and find these HiPS. Presently CDS maintains the HiPS directory, but there is clearly a role for virtual observatory registries to enhance the description and discovery of HiPS data sets.  

\section{Conclusions}
   
We have introduced HiPS as a practical multi-resolution HEALPix data structure for astronomical images, catalogues, and three-dimensional data cubes. There are already many ($\sim$180) HiPS image data sets available for a diverse range of sky survey data over all wavebands, and with sky coverages that range from a few square degrees up to the full sky, demonstrating the applicability this hierarchical method to many areas of astronomy.

The development of HiPS has placed a strong emphasis on conserving the scientific properties of the data alongside both visualisation considerations and emphasis on the ease of implementation. The scientific robustness of HEALPix enables HiPS to be a framework for multi-resolution statistical analysis of astronomical data sets in that each level of a HiPS is a statistical summary of the information in the lower levels. Also, there is much potential for the development of astronomical analysis procedures that operate directly on the HiPS pixel and tile grids, benefiting from the computational libraries developed for all sky HEALPix maps, that may lead to new capabilities for analysing partial sky maps. In addition to the scientific benefits related to HEALPix, HiPS also provides a simple mechanism for directly linking to the original data.

 The extension of HiPS to three-dimensional data cubes provides methods to visualise large data cubes and to manage many cube data sets simultaneously. The HiPS cube examples presented here illustrate the use of HiPS for cubes with a spectral axis as the third dimension, but this dimension may also be used to represent time or other kinds of axes. 
 
HiPS catalogues use the HiPS tile structure to organise astronomical source catalogues. We have shown that this method is a practical way of handling very large catalogues with our examples demonstrating good performance for catalogues with up to 2$\times$10$^9$ sources.  The development of HiPS catalogues is at a relatively early phase with the eleven examples presented here representing our experimentation with catalogues of different size, density, and for which we have tested different sorting keys. These techniques will be developed further as they are expected to be very important for future large catalogues that will be produced by GAIA, LOFAR, LSST, Euclid, and SKA and its pathfinders.
 
The combination of HiPS and MOC provides a very powerful tool for astronomy because they enable new levels of interoperability between astronomical images, data cubes, catalogues, and the description of regions of the sky. The simplicity and ease of use of the HiPS system is designed to facilitate wide spread implementation of HiPS across all areas of astronomy.

\begin{acknowledgements}
We thank William O'Mullane for useful discussions and comments on a near final version of this paper. We also acknowledge useful discussions facilitated by the IVOA and by the ADASS conference series.
Some of the results in this paper have been derived using the HEALPix \citep{2005ApJ...622..759G} package.
This research has made use of the SIMBAD database, the Aladin sky atlas, and the Vizier catalogue access tool  \citep{2000A&AS..143...23O} developed at CDS, Strasbourg Observatory, France.
We acknowledge the use of the Legacy Archive for Microwave Background Data Analysis (LAMBDA), part of the High Energy Astrophysics Science Archive Center (HEASARC). HEASARC/LAMBDA is a service of the Astrophysics Science Division at the NASA Goddard Space Flight Center.
This paper uses data from the CFHTLS. CFHTLS is based on observations obtained with MegaPrime/MegaCam, a joint project of CFHT and CEA/IRFU, at the Canada-France-Hawaii Telescope (CFHT) which is operated by the National Research Council (NRC) of Canada, the Institut National des Science de l'Univers of the Centre National de la Recherche Scientifique (CNRS) of France, and the University of Hawaii. This work is based in part on data products produced at Terapix available at the Canadian Astronomy Data Centre as part of the Canada-France-Hawaii Telescope Legacy Survey, a collaborative project of NRC and CNRS. This publication makes use of data products from the Wide-field Infrared Survey Explorer, which is a joint project of the University of California, Los Angeles, and the Jet Propulsion Laboratory/California Institute of Technology, funded by the National Aeronautics and Space Administration. 

\end{acknowledgements}

%-------------------------------------------------------------------

\bibliographystyle{aa}
\bibliography{Fernique_Allen_04May2015}

\clearpage

%_____________________________________________________________
%                                             Two column Table 
%_____________________________________________________________
%

\begin{table*}
\caption{HiPS tile and pixel properties for $n_{tile}$=512}            
\label{table:1}      
\centering          
\begin{tabular}{r r r r r r r}     
\hline\hline       
$k$   & $N_{side}=2^{k}$  & $N_{pix}$ & $\theta_{pix} $ & $k_{tile,512}$ & $N_{tile, 512}$ & $\theta_{tile, 512}$ \\
 & & & & & & \\
 \hline
0	& 1	     & 12	            & 58\fdg 6         & &   &       \\
1	& 2	     & 48	            & 29\fdg 3         & &    &      \\ 
2	& 4	     & 192	            & 14\fdg 7         & &     &    \\
3	& 8	     & 768	            &   7\fdg 33       & &     &    \\
4	& 16	     & 3072	            &   3\fdg 66       & &     &    \\
5	& 32	     & 12,288	             &   1\fdg 83      & &  &   \\
6	& 64	     & 49,152	             & 55\farcm 0      & &  & \\
7	& 128    &	196,608	             & 27\farcm 5      & &  & \\
8	& 256    & 786,432	             & 13\farcm 7      & &  & \\
9	& 512    &	3,145,728             &  6\farcm 87       & 0 &12       &      58\fdg 6     \\       
10	& 1024  &	12,582,912	     & 3\farcm 44         & 1 &48       &      29\fdg 3     \\   
11	& 2048  &	50,331,648	     & 1\farcm 72         & 2 & 192       &     14\fdg 7      \\  
12	& 4096  &	201,326,592	     & 51\farcs 5         & 3 & 768       &      7\fdg 33     \\ 
13	& 8192  &	805,306,368	     & 25\farcs 8         & 4 & 3072       &     3\fdg 66     \\ 
14	& 2$^{14}$&	3.22 $\times 10^{9}$ &	12\farcs 9        & 5 & 12288       &    1\fdg 83    \\ 
15	& 2$^{15}$&	1.29 $\times 10^{10}$	 & 6\farcs 44     & 6 & 49152       &    55\farcm 0   \\  
16	& 2$^{16}$&	5.15 $\times 10^{10}$	 & 3\farcs 22     & 7 &196608       &   27\farcm 5    \\ 
17	& 2$^{17}$&	2.06 $\times 10^{11}$	 & 1\farcs 61     & 8 & 786432       &   13\farcm 7    \\ 
18    & 2$^{18}$& 8.25 $\times 10^{11}$      & 0\farcs 81   	  & 9 & 3,145,728    &    6\farcm 87  \\ 
19    & 2$^{19}$&  3.30 $\times 10^{12}$     & 0\farcs 40   	  & 10 &12,582,912	 &    3\farcm 44   \\  
20   & 2$^{20}$ & 1.32 $\times 10^{13}$      & 0\farcs 20  	  & 11 & 50,331,648	 &    1\farcm 72    \\ 
21   & 2$^{21}$ & 5.28 $\times 10^{13}$      & 0\farcs 10 	  & 12 & 201,326,592	 &    51\farcs 5    \\ 
22   & 2$^{22}$ & 2.11 $\times 10^{14}$ &   50.3 mas    	  & 13 & 805,306,368	 &    25\farcs 8      \\	
23   & 2$^{23}$ & 8.44 $\times 10^{14}$ &   25.1 mas    	  & 14 & 3.22 $\times 10^{9}$   &  12\farcs 9  \\	
24   & 2$^{24}$ & 3.38 $\times 10^{15}$ &  12.6 mas     	  & 15 & 1.29 $\times 10^{10}$  &   6\farcs 44 \\	
25   & 2$^{25}$ & 1.35 $\times  10^{16}$ &  6.29 mas    	  & 16 & 5.15 $\times 10^{10}$  &   3\farcs 22  \\	
26   & 2$^{26}$ &  5.40 $\times 10^{16}$ &  3.15 mas    	  & 17 & 2.06 $\times 10^{11}$  &  1\farcs 61  \\	
27   & 2$^{27}$ &  2.16 $\times 10^{17}$ &  1.57 mas    	  & 18 & 8.25 $\times 10^{11}$  &  0\farcs 81      \\	
28   & 2$^{28}$ &  8.65 $\times 10^{17}$ &  0.786 mas   	  & 19 & 3.30 $\times 10^{12}$  &  0\farcs 40      \\	
29	& $2^{29}$ & 3.46 $\times 10^{18}$ & 0.393 mas   	  & 20 & 1.32 $\times 10^{13}$  & 0\farcs 20      \\	 
\hline                  
\end{tabular}
\end{table*}

\begin{table*}
\caption{HiPS tile and pixel properties of single order HiPS with $n_{tile}$=64, 128, and 256}             
\label{table:2}      
\centering          
\begin{tabular}{r r r r r r r}     
\hline\hline  
\multicolumn{7}{c}{\em $n_{tile}$=64 } \\   
\hline  
$k$   & $N_{side}$  & $N_{pix}$ & $\theta_{pix} $ & $k_{tile,64}$ & $N_{tile, 64}$ & $\theta_{tile, 64}$ \\
\hline
9	& 512    &	3,145,728            &  6\farcm 87      & 3 & 768       &      7\fdg 33     \\ 			  	 
        &                &                                         &                       &        &                    &  \\	  
\hline
\hline
\multicolumn{7}{c}{\em $n_{tile}$=128 } \\   
\hline  
$k$   & $N_{side}$  & $N_{pix}$ & $\theta_{pix} $ & $k_{tile,128}$ & $N_{tile, 128}$ & $\theta_{tile, 128}$ \\
\hline
10	& 1024  &	12,582,912	     & 3\farcm 44       & 3 & 768       &      7\fdg 33     \\ 
        &                &                                         &                       &        &                    &  \\	  
			  	 
\hline
\hline
\multicolumn{7}{c}{\em $n_{tile}$=256 } \\   
\hline  
$k$   & $N_{side}$  & $N_{pix}$ & $\theta_{pix} $ & $k_{tile,256}$ & $N_{tile, 256}$ & $\theta_{tile, 256}$ \\
\hline
11	& 2048  &	50,331,648	     & 1\farcm 72       & 3 & 768       &      7\fdg 33     \\ 			  
        &                &                                         &                       &        &                    &  \\	  
\hline              
\end{tabular}
\end{table*}

\clearpage

%\begin{comment}
\begin{longtab}
\begin{longtable}{l l r c r }    
\caption{HiPS images \label{table:HipsImage}} \\
\hline
\hline
ID\tablefootmark{a} & Data set name and description & $k_{tile}$ & $\theta_{pix}$ & Sky   \\
     &                                 &                &                       &    (\%)    \\      
\hline
\endfirsthead
\multicolumn{5}{c}{ } \\
\multicolumn{5}{c}{\em{-- Gamma Ray --}} \\
\multicolumn{5}{c}{ } \\
%\hline
 P/Fermi\tablefootmark{E}\tablefootmark{b}\tablefootmark{c}\tablefootmark{d} \dotfill& Fermi$^{(1)}$: 4 HiPS in 3-300\ MeV, 300-1000\ MeV, & 3 & 6\farcm87 & 100\%\\
               &    and 1-3\ GeV and a colour composite ($n_{tile}$=64)                                                   &   &           &            \\      
P/EGRET\tablefootmark{E}\tablefootmark{b}\tablefootmark{c}\tablefootmark{d}  \dotfill & EGRET$^{(2)}$:  12 HiPS for 100 MeV {\em inf} and {\em sup} maps, and & 3 & 51\farcs5 & 100\% \\
                          & diffuse maps 30-50, 50-70, 70-100, 100-150, 150-300, &     &                &           \\
                          & 300-500, 500-1000 MeV, 1-2 GeV, 2-4 GeV, 4-10 GeV &     &                &           \\
\hline
\multicolumn{5}{c}{ } \\
\multicolumn{5}{c}{{\em -- X-ray --}} \\
\multicolumn{5}{c}{ } \\
%\hline
 P/XMM\tablefootmark{E}\tablefootmark{e}\tablefootmark{f} \dotfill & XMM-Newton : 5 HiPS for EPIC stacked images, PN   & 7 & 3\farcs22 & 7\%\\
               &  in 0.5-1, 1-2, and 2-4.5 keV bands and PN colour      &     &                &            \\     
 P/INTEGRAL\tablefootmark{E}\tablefootmark{g}\tablefootmark{c}\tablefootmark{d} \dotfill & INTEGRAL$^{(3)}$  SPI GC (20-40 keV) & 3 & 51\farcs5 & 17\% \\
 P/INTGAL\tablefootmark{E}\tablefootmark{c}\tablefootmark{d}\tablefootmark{h}  \dotfill & INTEGRAL IBIS$^{(4)}$: 4 HiPS for 17-35, 17-60, & 3 & 51\farcs5 & 31\% \\
               &     35-80 keV, and colour composition                             &                &                       &     \\     
 P/RASS\tablefootmark{G}\tablefootmark{c}\tablefootmark{d}\tablefootmark{i} \dotfill & ROSAT X-Ray All Sky Survey$^{(5)}$ & 4 & 25\farcs8 & 100\%\\
\hline
\multicolumn{5}{c}{ } \\
\multicolumn{5}{c}{\em -- Ultraviolet --} \\
\multicolumn{5}{c}{ } \\
%\hline  
 P/GALEXGR6\tablefootmark{E}\tablefootmark{c}\tablefootmark{d}\tablefootmark{j} \dotfill & GALEX$^{(6)}$ :  Colour composition of GR6 AIS  & 8 & 1\farcs61 & 69\%\\
\hline
\multicolumn{5}{c}{ } \\
\multicolumn{5}{c}{\em -- Optical --} \\
\multicolumn{5}{c}{ } \\
%\hline
 P/DSS2\tablefootmark{E}\tablefootmark{c}\tablefootmark{d}\tablefootmark{j}  \dotfill & DSS2: 2 HiPS for red band and colour composite & 9 & 0\farcs81 & 100\%\\
  %    &                                 &                &                       &      \\    
 P/SDSS9\tablefootmark{E}\tablefootmark{c}\tablefootmark{d}\tablefootmark{k} \dotfill & SDSS DR9$^{(7)}$: 3 HiPS for r, g, and a colour composite & 10 & 0\farcs4 & 36\%\\
%     &                                 &                &                       &      \\    
 P/Mellinger/color\tablefootmark{E}\tablefootmark{c}\tablefootmark{d}\tablefootmark{m} \dotfill & Mellinger optical survey & 4 & 25\farcs77 & 100\%\\
 %    &                                 &                &                       &      \\    
 P/CFHTLS/D\tablefootmark{E}\tablefootmark{c}\tablefootmark{d}\tablefootmark{n} \dotfill & CFHTLS$^{(8)}$ Deep : 6 HiPS for u,g,r,i,z, and colour & 12 & 0\farcs1 &   0.01\%      \\
  P/CFHTLS/W\tablefootmark{E}\tablefootmark{c}\tablefootmark{d}\tablefootmark{n} \dotfill & CFHTLS$^{(8)}$ Wide : 6 HiPS for u,g,r,i,z, and colour & 11 & 0\farcs2 & 0.4\%\\
  %   &                                 &                &                       &      \\    
 P/HST\tablefootmark{E}\tablefootmark{d}\tablefootmark{p}\tablefootmark{q} \dotfill & HST: 17 HiPS for NICMOS, WFPC2, and ACS in    & 14 & 25.1\ mas & 0.044\%\\
                       &  F110W, F160W, F225W, F300W, F450W, F606W, &     &          &   \\
                        & F625W, F702W, F814W, F850LP,  and  GOODS$^{(9)}$   &      &         &      \\    
                      &   associations in b,v,i,z, and colour            &       &         &      \\    
% pagebreak here for referee mode
%\pagebreak
 \hline
 \multicolumn{5}{c}{ } \\
\multicolumn{5}{c}{\em -- Infrared --} \\
%\hline
\multicolumn{5}{c}{ } \\
P/2MASS\tablefootmark{E}\tablefootmark{c}\tablefootmark{d}\tablefootmark{r} \dotfill  & 2MASS$^{(10)}$: 4 HiPS for J, H, K, and colour & 9 & 0\farcs81 & 100\%\\
P/2MASS6X\tablefootmark{E}\tablefootmark{c}\tablefootmark{d}\tablefootmark{r} \dotfill  & 2MASS6X: 3 HiPS for J, H, and K  & 9 & 0\farcs81 & 1.32\% \\
 P/ULTRAVISTA\tablefootmark{E}\tablefootmark{c}\tablefootmark{d}\tablefootmark{s}  \dotfill & VISTA Ultra Deep Survey$^{(11)}$: 6 HiPS for Y, J,  & 12 & 0\farcs1 & 0.005\%\\
                            & H, Ks, NB118, and colour                             &     &         &   \\ 
 P/WISE\tablefootmark{E}\tablefootmark{c}\tablefootmark{d}\tablefootmark{r} \dotfill & WISE$^{(12)}$: 5 HiPS for 3.4, 4.6, 12, 22 $\mu$m, and colour   &  5 & 12\farcs9 & 100\%\\
 P/WISE/WSSA\tablefootmark{E}\tablefootmark{c}\tablefootmark{d} \dotfill & WSSA$^{(13)}$ Diffuse dust emission at 12 $\mu$m & 7 & 3\farcs22 & 100\% \\
 P/ALLWISE\tablefootmark{E}\tablefootmark{c}\tablefootmark{d}\tablefootmark{r} \dotfill & ALLWISE$^{(14)}$: 5 HiPS for 3.4, 4.6, 12, 22 $\mu$m, and colour   &  8 & 1\farcs61 & 100\%\\
 P/DIRBE\tablefootmark{G}\tablefootmark{u}\tablefootmark{v} \dotfill & DIRBE$^{(15)}$: 20 HiPS for bands 1-10 and ZSMA bands 1-10 & 3 & 51\farcs5 & 100\%\\
%                          &                              &     &         &   \\ 
 %
 P/IRIS\tablefootmark{G}\tablefootmark{u}\tablefootmark{v}\tablefootmark{r} \dotfill  & IRAS-IRIS$^{(16)}$: 4 HiPS for 12, 25, 60, and 100 $\mu$m  & 3 & 51\farcs5 & 100\%\\
 P/IRIS\tablefootmark{G}\tablefootmark{c}\tablefootmark{d}\tablefootmark{r} \dotfill  & IRAS-IRIS$^{(16)}$: HiPS colour composition  & 3 & 51\farcs5 & 100\%\\
 P/GLIMPSE360\tablefootmark{E}\tablefootmark{r}\tablefootmark{w}\tablefootmark{x}\tablefootmark{z}  \dotfill & GLIMPSE 360$^{(17,18)}$: colour & 9 & 0\farcs81 & 3.1\%\\
 P/SPITZER/IRAC\tablefootmark{c}\tablefootmark{d}\tablefootmark{y}\tablefootmark{x} \dotfill  &  Spitzer IRAC: 5 HiPS for the 3.6, 4.5, 5.8,   & 9 & 0\farcs81 & 1.4\%\\
                               &  8$\mu$m, and colour bands                                       &    &                 & \\
 P/SPITZER/MIPS1\tablefootmark{G}\tablefootmark{c}\tablefootmark{d}\tablefootmark{y}\tablefootmark{x} \dotfill & Spitzer MIPS1$^{(22)}$ 24 $\mu$m band & 8 & 1\farcs61 & 1.6\% \\
 P/SPITZER/MIPS2\tablefootmark{G}\tablefootmark{c}\tablefootmark{d}\tablefootmark{y}\tablefootmark{x} \dotfill & Spitzer MIPS2$^{(22)}$ 70 $\mu$m band & 7 & 3\farcs22 & 1.6\% \\
 P/SPITZER/MIPS3\tablefootmark{G}\tablefootmark{c}\tablefootmark{d}\tablefootmark{y}\tablefootmark{x} \dotfill & Spitzer MIPS3$^{(22)}$ 160 $\mu$m band & 6 & 6\farcs44 & 1.6\% \\
 P/AKARI/FIS\tablefootmark{E}\tablefootmark{c}\tablefootmark{d}\tablefootmark{$\alpha$}  \dotfill & AKARI$^{(23)}$ Far infrared all-sky survey: 5 HiPS  for N60,  & 5  & 12\farcs9&  99.9\%\\ 
                                  &N160, WideS, WideL, and colour  &    &  & \\
% pagebreak here for formatting in one column mode
\pagebreak
\hline
\multicolumn{5}{c}{ } \\
\multicolumn{5}{c}{\em -- Sub-mm and radio --} \\
\multicolumn{5}{c}{ } \\
P/SCUBA/450em\tablefootmark{E}\tablefootmark{c}\tablefootmark{d}\tablefootmark{$\beta$} \dotfill  & SCUBA 450$^{(24)}$ $\mu$m  & 8 & 1\farcs61 & 0.3\%\\
 P/SCUBA/850em\tablefootmark{E}\tablefootmark{c}\tablefootmark{d}\tablefootmark{$\beta$} \dotfill  & SCUBA 850$^{(24)}$ $\mu$m  & 7 & 3\farcs22 & 0.9\%\\
 P/SCUBA/850emi\tablefootmark{E}\tablefootmark{c}\tablefootmark{d}\tablefootmark{$\beta$} \dotfill & SCUBA 850$^{(24)}$ $\mu$m extended data set & 7 & 3\farcs22 & 0.9\%\\
 P/SCUBA2\tablefootmark{E}\tablefootmark{d}\tablefootmark{$\gamma$} \dotfill & SCUBA-2$^{(25)}$  : 2 HiPS for 450 and 850 $\mu$m & 9 & 0\farcs81 & 1.8\% \\
 P/BOLOCAM\tablefootmark{G}\tablefootmark{u}\tablefootmark{v} \dotfill & Bolocam Galactic plane survey$^{(26)}$ (BGPS) & 5 & 12\farcs9 & 2\%\\
 P/WMAP\tablefootmark{G}\tablefootmark{c}\tablefootmark{d}\tablefootmark{$\delta$} \dotfill & WMAP Year 9 results$^{(27)}$: 5 HiPS for W, V, Q, Ka, and K & 3 & 6\farcm87 & 100\%\\
                  & ($n_{tile}$=64) & & & \\
P/PLANCK/HFI\tablefootmark{G}\tablefootmark{c}\tablefootmark{d}\tablefootmark{$\epsilon$}  \dotfill & PLANCK$^{(28)}$ (2013 data release): 6 HiPS for HFI in HFI100, & 3 & 1\farcm72 & 100\%\\
                                      &  143, 217, 353, 545, 857 GHz, and colour  ($n_{tile}$=256)  & & &  \\
P/PLANCK/LFI\tablefootmark{G}\tablefootmark{c}\tablefootmark{d}\tablefootmark{$\epsilon$}  \dotfill & PLANCK$^{(28)}$ (2013 data release): 4 HiPS for LFI in 1030,  & 3 & 3\farcm44 & 100\% \\
                                    & 1044, 1070 GHz, and colour  ($n_{tile}$=256)                                    &    &                &       \\
P/PLANCK/CMB\tablefootmark{G}\tablefootmark{c}\tablefootmark{d}\tablefootmark{$\epsilon$}    \dotfill & PLANCK CMB fluctuations map (SMICA inpainted)   & 3 & 1\farcm72 & 100\% \\
                                   & ($n_{tile}$=256) & & & \\
  P/NVSS\tablefootmark{E}\tablefootmark{c}\tablefootmark{d}\tablefootmark{$\zeta$} \dotfill & The NRAO VLA Sky Survey$^{(29)}$  & 5 & 12.88 & 84.8\%\\
  P/CHIPASS\tablefootmark{G}\tablefootmark{u}\tablefootmark{v} \dotfill & CHIPASS$^{(30)}$ 1.4 GHz Radio continuum survey & 3 & 51\farcs5 & 72\%  \\
 P/SUMSS\tablefootmark{E}\tablefootmark{c}\tablefootmark{d} \dotfill & Sydney University Molonglo sky survey$^{(31,32)}$ 843MHz & 6 & 6\farcs44 & 25\% \\ 
 P/DWINGELOO\tablefootmark{G}\tablefootmark{u}\tablefootmark{v}  \dotfill & Dwingeloo 820 MHz continuum survey$^{(33)}$ & 3 & 51\farcs5 & 57\% \\
 P/HASLAM408\tablefootmark{G}\tablefootmark{$\eta$}\tablefootmark{$\theta$}\tablefootmark{d} \dotfill & Haslam 408 MHz survey: 2 HiPS for original$^{(34)}$  & 3 & 51\farcs5 & 100\% \\
           & and reprocessed$^{(35)}$ data & & & \\
P/VLSSR\tablefootmark{E}\tablefootmark{c}\tablefootmark{d}\tablefootmark{$\zeta$} \dotfill & VLA low-frequency sky survey redux$^{(36)}$  & 5 & 12\farcs9 & 75\% \\
           & (74 MHz continuum) & & & \\
 P/WENSS\tablefootmark{E}\tablefootmark{c}\tablefootmark{d}\tablefootmark{$\kappa$} \dotfill & Westerbork northern sky survey$^{(37)}$ (325 MHz)& 5 & 12.88 & 26.9\%\\
\hline
\multicolumn{5}{c}{ } \\
\multicolumn{5}{c}{\em -- H$\alpha$ --}  \\
\multicolumn{5}{c}{ } \\
 P/Finkbeiner\tablefootmark{G}\tablefootmark{c}\tablefootmark{d}\tablefootmark{$\lambda$}  \dotfill & Finkbeiner H$\alpha$ composite survey$^{(38)}$ & 3 & 51\farcs5 & 100\%\\
 P/SHS\tablefootmark{E}\tablefootmark{c}\tablefootmark{d}\tablefootmark{$\mu$} \dotfill & SuperCOSMOS digitized photographic H$\alpha$ Survey$^{(39)}$  & 10 & 0\farcs4 & 9.8\%\\
 P/SHASSA\tablefootmark{G}\tablefootmark{c}\tablefootmark{d}\tablefootmark{$\xi$} \dotfill & SHASSA southern H$\alpha$ sky survey atlas$^{(40)}$:  4 HiPS for          & 4 & 25\farcs77 & 67\%\\
                              &   H$\alpha$, continuum, continuum-corrected,  H$\alpha$     &  & &  \\
                             & and smoothed H$\alpha$ images                     &   &                 &   \\

 P/VTSS\tablefootmark{G}\tablefootmark{c}\tablefootmark{d}\tablefootmark{$\pi$} \dotfill & Virginia Tech spectral line survey$^{(41)}$: 2 HiPS for & 3 & 51\farcs5 & 20\% \\
                         & H$\alpha$ and continuum corrected H$\alpha$ &   &                 &  \\
 \hline
 \multicolumn{5}{c}{ } \\
\multicolumn{5}{c}{\em -- HI --} \\
\multicolumn{5}{c}{ } \\
 P/GASS\tablefootmark{G}\tablefootmark{c}\tablefootmark{d}\tablefootmark{$\rho$}  \dotfill & Parkes Galactic all sky survey$^{(42)}$ (GASS) ($n_{tile}$=64) & 3 & 6\farcm87 & 53\%\\
 P/GASS/NH\tablefootmark{G}\tablefootmark{u}\tablefootmark{v}\tablefootmark{$\rho$} \dotfill & Parkes Galactic all sky survey$^{(42)}$ (GASS) ($n_{tile}$=128) & 3 & 3\farcm44 & 53\% \\
 P/CGPS/VGPS\tablefootmark{G}\tablefootmark{u}\tablefootmark{v}\tablefootmark{$\sigma$}\tablefootmark{$\tau$} \dotfill & Combination CGPS$^{(43)}$ and VGPS$^{(44)}$: 2 HiPS for  & 4 & 25\farcs8 & 3.9\%\\
                                     & 1420 MHz continuum and HI                         &   &            &        \\
\hline
\multicolumn{5}{c}{ } \\
\multicolumn{5}{c}{\em -- CO --} \\
\multicolumn{5}{c}{ } \\
 P/CO\tablefootmark{G}\tablefootmark{d}\tablefootmark{$\delta$} \dotfill & CO composite survey$^{(45)}$ & 3 & 51\farcs5 & 100\%\\
 P/COHRS\tablefootmark{E}\tablefootmark{c}\tablefootmark{d}\tablefootmark{$\delta$} \dotfill & CO high resolution survey$^{(46)}$ (HARP/JCMT) & 7 & 3\farcs22 & 0.15\%\\
\hline
\end{longtable}
\tablefoot{
\tablefoottext{E}{Equatorial (ICRS) coordinate system HiPS};
\tablefoottext{G}{Galactic coordinate system HiPS};
\tablefoottext{a}{CDS identifier for the HiPS data set};
\tablefoottext{b}{Original data distributed by SkyView/HEASARC};
\tablefoottext{c}{HEALPixed by CDS};
\tablefoottext{d}{HiPS distributed by CDS};
\tablefoottext{e}{HEALPixed by Laurent Michel (Observatoire de Strasbourg)};
\tablefoottext{f}{HiPS distributed by XMM-Newton science survey consortium, EC FP7 ARCHES project};
\tablefoottext{g}{Original data distributed by INTEGRAL Science Data Centre, Geneva};
\tablefoottext{h}{Original data distributed by Space Research Institute (IKI), Moscow};
\tablefoottext{i}{Original data distributed by Max Planck Institut fur extraterrestrische Physik};
\tablefoottext{j}{Original data distributed by MAST at STScI};
\tablefoottext{k}{Original data distributed by the SLOAN Digital Sky Survey};
\tablefoottext{m}{Original data (c) Axel Mellinger};
\tablefoottext{n}{Original data (c) CFH - powered by Terapix};
\tablefoottext{p}{Processed data from CADC};
\tablefoottext{q}{HEALPixed by Daniel Durand (CADC)};
\tablefoottext{r}{Original data distributed by IPAC/NASA Infrared Science Archive (IRSA)};
\tablefoottext{s}{Original data distributed by the UltraVISTA consortium};
\tablefoottext{t}{Processed data obtained from (12)};
\tablefoottext{u}{HiPS distributed by Analysis Centre for Extended Data (CADE), IRAP Toulouse};
\tablefoottext{v}{HEALPixed by D. Paradis, CADE, IRAP Toulouse};
\tablefoottext{w}{HEALPixed by T. Robitaille};
\tablefoottext{x}{Original data from Spitzer mission, JPL/NASA};
\tablefoottext{y}{Composite from Spitzer legacy programs: GLIMPSE (17,18), SAGE (19), SAGE-SMC (20) and SINGS (21)};
\tablefoottext{z}{HiPS distributed by Spitzer mission, JPL/NASA};
\tablefoottext{$\alpha$}{Original data distributed by ISAS, JAXA, Japan};
\tablefoottext{$\beta$}{Processed data obtained from CADC};
\tablefoottext{$\gamma$}{HiPS built from JCMT data via CADC archive};
\tablefoottext{$\delta$}{Original data from LAMBDA};
\tablefoottext{$\epsilon$}{Original data from PLANCK mission at ESA};
\tablefoottext{$\zeta$}{Original data from NRAO};
\tablefoottext{$\eta$}{Data distributed and maintained by LAMBDA, who cite Jodrell Bank Centre for Astrophyscis as the original data source};
\tablefoottext{$\theta$}{Reprocessed data by described in (35) including details of various versions of the 408 MHz Haslam map};
\tablefoottext{$\kappa$}{Original data from WENSS at ASTRON in the Netherlands};
\tablefoottext{$\lambda$}{Composed by VTSS (Virginia Tech spectral line survey), SHASSA (Southern H$\alpha$ sky survey atlas), WHAM (Wisconsin H$\alpha$ mapper)};
\tablefoottext{$\mu$}{Original data courtesy of WFAU, AAO/UKST, PPARC/STFC};
\tablefoottext{$\xi$}{Original data from Swarthmore College Incorporated and available from NASA/Skyview};
\tablefoottext{$\pi$}{Original data from Virginia Tech Physics};
\tablefoottext{$\rho$}{Original data from CSIRO/ATNF};
\tablefoottext{$\sigma$}{Original VGPS data from University of Calgary and CADC};
\tablefoottext{$\tau$}{Original CGPS data from Canadian Galactic plane consortium and available from CADC}
\\ }
\tablebib{
(1) \citet{2009ApJ...697.1071A};
(2) \citet{2005ApJ...621..291C};
(3) \citet{2003A&A...411L...1W};
(4) \citet{2012A&A...545A..27K};
(5) \citet{1992ESOC...43..139V};
(6) \citet{2005ApJ...619L...1M};
(7) \citet{2012ApJS..203...21A};
(8) see CFHTLS in the acknowledgements;
(9) \citet{2004ApJ...600L..93G};
(10) \citet{2006AJ....131.1163S};
(11) \citet{2012A&A...544A.156M};
(12) \citet{2010AJ....140.1868W};
(13) \citet{2014ApJ...781....5M};
(14) \citet{2013wise.rept....1C};
(15) \citet{2005ApJS..157..302M};
(16) \citet{1992ApJ...397..420B};
(17) \citet{2003PASP..115..953B};
(18) \citet{2009PASP..121..213C};
(19) \citet{2008ASPC..381..115M};
(20) \citet{2011AJ....142..102G};
(21) \citet{2003PASP..115..928K};
(22) \citet{2004ApJS..154....1W};
(23) \citet{2007PASJ...59S.369M};
(24) \citet{1999MNRAS.303..659H};
(25) \citet{2013MNRAS.430.2513H};
(26) \citet{2011ApJS..192....4A};
(27) \citet{2013ApJS..208...20B};
(28) \citet{2014A&A...571A...1P};
(29) \citet{1998AJ....115.1693C};
(30) \citet{2014PASA...31....7C};
(31) \citet{2003MNRAS.342.1117M};
(32) \citet{2007MNRAS.382..382M};
(33) \citet{1972A&AS....5..263B};
(34) \citet{1982A&AS...47....1H};
(35) \citet{2014arXiv1411.3628R} 
(36) \citet{2012RaSc...47.0K04L};
(37) \citet{1997A&AS..124..259R};
(38) \citet{2003ApJS..146..407F};
(39) \citet{2005MNRAS.362..689P};
(40) \citet{2001PASP..113.1326G};
(41) \citet{1999AAS...195.5309D};
(42) \citet{2009ApJS..181..398M};
(43) \citep{2003AJ....125.3145T};
(44) \citet{2006AJ....132.1158S}; 
(45) \citet{2001ApJ...547..792D};
(46) \citet{2013ApJS..209....8D};
(47) \citet{2008ApJS..175..277D} 
}
\end{longtab}
%\end{comment}

\clearpage

\setcounter{table}{3}

\begin{table*}
\caption{HiPS catalogues}             
%\label{table:HipsCat}      
\centering          
\begin{tabular}{llrlr}     
\hline\hline
ID\tablefootmark{a} & Data set name and description & $k_{tile,max}$ & Sort-key\tablefootmark{b} & $N(sources)$  \\
                                &                                                   &                        &               &                    \\      
\hline
B/SIMBAD\dotfill                & CDS SIMBAD database$^{(1)}$\tablefootmark{c}              & 14            & Literature citations   & $\sim$8$\times10^{6} $ \\
I/259/TYC2        \dotfill       & Tycho-2 catalogue$^{(2,3,4)}$                                          &  6             &  BT,VT                                  & 2,539,913                        \\
I/312             \dotfill           & PPMX catalogue of positions and proper motions$^{(5,6)}$   &  9             &   B1,B2,R1,Rs,I                  & 18,088,919                       \\
I/324             \dotfill            & Initial Gaia source list$^{(7,8)}$                                         &  11           &    BJ,RF,G,Grvs                  & 1,222,598,530                  \\
II/246              \dotfill           & 2MASS$^{(9,10)}$                                                             & 11            &   J,H,K                                 &  470,992,970                    \\
II/293               \dotfill          & GLIMPSE$^{(11,12,13)}$                                                   & 10            &   f3.6,4.5,5.8,8.0 $\mu$m   & 104,240,613                     \\
II/294              \dotfill          & SDSS-DR7$^{(14)}$                                                            &  9             &   u,g,r,i,z                             & 65,714,108                      \\
II/297             \dotfill           & AKARI IRC$^{(15,16)}$                                                       & 7              &    S09                                & 870,973                          \\
II/298             \dotfill           & AKARI FIS$^{(17)}$                                                            & 6              &     S65                                     & 427,071                          \\
II/311/WISE \dotfill             & WISE$^{(18,19)}$                                                               & 8              &   W1 band                           & 563,921,584                   \\
VI/137             \dotfill          & Gaia universe model snapshot$^{(20,21)}$ (GUMS)        & 11            & Distance                             &  2,143,475,885               \\      
\hline
\end{tabular}
\tablefoot{
\tablefoottext{a}{CDS identifier for the catalogue};
\tablefoottext{b}{The sort-key used to distribute catalogue sources over the different orders of the HiPS. Where the sort-key is based on source brightness, we indicate the list of the bandpasses that were combined to define the sort-key};
\tablefoottext{c}{SIMBAD is a living database that is continually updated with new astronomical objects, hence $N(sources)$ is growing.}
}
\tablebib{
(1) \citet{2000A&AS..143....9W};
(2) \citet{2000A&A...357..367H};
(3) \citet{2000A&A...355L..27H};
(4)  \citet{2000yCat.1259....0H};
(5) \citet{2008A&A...488..401R};
(6) \citet{2008yCat.1312....0R};
(7) \citet{2014A&A...570A..87S};
(8)  \citet{2013yCat.1324....0S};
(9) \citet{2003tmc..book.....C};
(10) \citet{2003yCat.2246....0C};
(11) \citet{2009PASP..121..213C};
(12) \citet{2003PASP..115..953B};
(13) \citet{2009yCat.2293....0S};
(14) \citet{2009ApJS..182..543A};
(15) \citet{2010A&A...514A...1I};
(16) \citet{2010yCat.2297....0I};
(17) \citet{2010yCat..35149003P};
(18) \citet{2010yCat..35149003P};
(19) \citet{2010AJ....140.1868W};
(20) \citet{2012A&A...543A.100R};
(21) \citet{2012yCat..35439100R}
}
\end{table*}

%\clearpage

\begin{table*}
\caption{HiPS three-dimensional data cubes}                   
\centering          
\begin{tabular}{llrcrr}     
\hline\hline
ID\tablefootmark{a}   & Data set name and description & $k_{tile,max}$  & $\theta_{pix,min}$ & $N(planes)$ & Sky  \\
                                  &                                                   &                         &                               &                      &      (\%)    \\      
\hline
C/EGRET/DIF \dotfill & EGRET Diffuse Maps$^{(3)}$   & 3                      & 51\farcs5               & 10                 & 100\%      \\
                                  & 30-50, 50-70, 70-100, 100-150,   &                      &                               &                      &                  \\
                                  & 150-300, 300-500, 500-1000 MeV, &                   &                               &                      &                  \\
                                  & 1-2 GeV, 2-4 GeV, and 4-10 GeV &                            &                               &                      &                  \\
C/HST/GOODS\tablefootmark{b}\tablefootmark{c}   \dotfill    & GOODS$^{(4)}$  project HST data  $(b,v,i,z)$        & 14       & 25.1 mas  & 4              &                        \\
C/HST\tablefootmark{b}\tablefootmark{c}      \dotfill               & HST:  NICMOS, WFPC2, and ACS in                       & 14       & 25.1\ mas & 13            &   0.044\%        \\
                                 &  F110W, F160W, F225W, F300W,                           &            &                   &                &                 \\ 
                                 & F450W, F606W, F625W, F702W,                            &            &                   &                &                 \\ 
                                 & F814W, and F850LP                                                       &            &                   &                &                 \\ 
C/CALIFA/V500/DR1 \dotfill & CALIFA DR1 V500 cubes$^{(6)}$                         & 9        & 0\farcs81     & 1877      &     0.00009\%              \\
C/CALIFA/V500/DR2 \dotfill & CALIFA DR2 V500 cubes$^{(7)}$                         & 9        & 0\farcs81     & 1877      &      0.00018\%             \\
C/ALLWISE    \dotfill          & ALLWISE$^{(2)}$   3.4, 4.6, 12, and 22 $\mu$m           & 8         & 1\farcs61     & 4            & 100\%        \\
C/IRIS      \dotfill               & IRAS-IRIS$^{(5)}$   12, 25, 60, and 100 $\mu$m    & 3         & 51\farcs5    & 4             & 100\%        \\
C/CGPS  \dotfill                 & Canadian Galactic plane survey$^{(1)}$               & 5         &  12\farcs9    &  272       &  3.9\%         \\
C/WMAP/CUBE/9YR \dotfill & WMAP W, V, Q, Ka, and K ($n_{tile}$=64)           & 3         & 6\farcm87    & 5            &  100\%       \\
\hline
\end{tabular}
\tablefoot{
\tablefoottext{a}{CDS identifier for the HiPS cube};
\tablefoottext{b}{Processed data from CADC};
\tablefoottext{c}{HEALPixed by Daniel Durand (CADC)}
}
\tablebib{
(1) \citet{2003AJ....125.3145T};
(2) \citet{2013wise.rept....1C};
(3) \citet{2005ApJ...621..291C};
(4) \citet{2004ApJ...600L..93G};
(5) \citet{1992ApJ...397..420B};
(6) \citet{2013A&A...549A..87H};
(7) \citet{GarciaBenito:2014wg}
}
\end{table*}

\end{document}